\documentclass[aip,
amsmath,amssymb,
preprint,
11pt,
]{revtex4-2}

\usepackage{graphicx}% Include figure files
\usepackage{dcolumn}% Align table columns on decimal point
\usepackage{bm}% bold math
\usepackage{physics}

\usepackage[T1]{fontenc}
\usepackage{lmodern}

\usepackage{multirow}
\usepackage{xcolor}
\usepackage{hyperref}
\hypersetup{
    colorlinks=true,
    linkcolor=blue,
    filecolor=blue,      
    urlcolor=blue,
    citecolor=blue,
}

\begin{document}

\preprint{AIP/JoR/preprint}

\title{Thixotropic spectra and Ashby-style charts for thixotropy}

\author{Samya Sen}
\author{Randy H. Ewoldt}
 \email{ewoldt@illinois.edu}
\affiliation{Department of Mechanical Science and Engineering,\\University of Illinois at Urbana-Champaign,\\Urbana, IL 61801, USA}

\date{\today}

\begin{abstract}
There is no universal model for thixotropy, and comparing thixotropic effects between different fluids is a subtle yet challenging problem. We introduce a generalized (model-insensitive) framework for comparing thixotropic properties based on thixotropic spectra. A superposition of exponential stress modes distributed over thixotropic timescales is used to quantify buildup and breakdown times and mode strengths in response to step-change input. This mathematical framework is tested with several experimental step-shear rate data on colloidal suspensions. Low-dimensional metrics based on moments of the distribution reveal characteristic average thixotropic properties which are visualized on Ashby-style diagrams. This method outlines a framework for describing thixotropy across a diverse range of microstructures, supporting scientific studies as well as material selection for engineering design applications.
\end{abstract}

\maketitle

%%%%%%%%%%%%%%%%%%%%%%%%%%%%%%%%%%%%%%%%%%%%%%%%%%%%%%%%%%%%%%%%%%%%%%%%%%%%%%%%
%===============================================================================
% section break
%===============================================================================
%%%%%%%%%%%%%%%%%%%%%%%%%%%%%%%%%%%%%%%%%%%%%%%%%%%%%%%%%%%%%%%%%%%%%%%%%%%%%%%%

\section{Introduction\label{sec:intro}}

Thixotropic phenomena are observed in fluids with a ``structure'' that evolves with time, and such fluids are common both in industry and in daily life. Common examples include flow batteries \cite{Helal_2014,Helal_2016,Narayanan,Wang_JoR}, crude oils \cite{DimitriouMcKinley_SM2014}, food materials \cite{Glicerina2015}, blood \cite{Jin_Blood2011,Armstrong2020}, and colloidal suspensions \cite{DullaertMewis_ModelThixo2005,Alessandrini1982,Beris_starJNNFM2008,Burgos2001,Kelessidis2008}, among numerous others \cite{MewisReview1979,Barnes_ThixoReview1997,MewisWagnerReview2009,LarsonWei-JoR2019}. As defined by Mewis and Wagner \cite{MewisWagnerReview2009,MewisWagner_book2012}, thixotropy is ``the continuous decrease of viscosity with time when flow is applied to a sample that has been previously at rest, and the subsequent recovery of viscosity when the flow is discontinued''. Real materials show both thixotropy and viscoelasticity, and microstructural buildup can cause an increase in the viscosity as well as the elastic modulus \cite{Joshi_JOR2021}. In many cases thixotropy may be distinguished from viscoelasticity based on timescales: viscoelastic timescales are typically shorter compared to the thixotropic timescales \cite{WeiSolomonLarson_SE-JOR2016,Larson_ConstEq-JOR2015}.

% subsection break -----------------------------------------------------

\begin{figure}[!ht]
	\centering
	\includegraphics[width=0.8\linewidth,trim={0 0 5cm 0},clip]{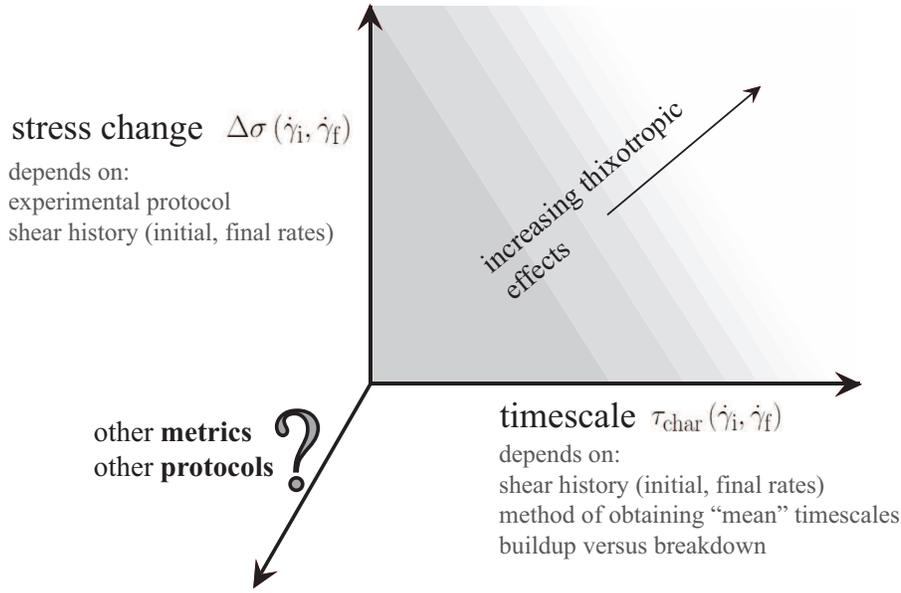}
	\caption{\label{fig:intro_Ashby}Proposed Ashby-style mapping for comparing thixotropic responses. Even with a chosen experimental protocol, the dependencies of two relevant thixotropic properties on the experimental conditions makes the parameter space high-dimensional.}
\end{figure}

Materials with short thixotropic timescales (e.g.\ $\lesssim\mathcal{O}(1)$~s) are typically labeled as \emph{not} thixotropic, but the degree of thixotropy of a material depends on not only the characteristic timescales, but also the amount of change in a rheological property of interest, e.g.\ shear viscosity or elastic modulus, or more generally a change in state of stress. Stress changes, and timescales over which these changes occur, must both be considered to quantify and compare the degree of thixotropy across different materials. Additionally, thixotropic materials involve a range of microstructural lengthscales and associated timescales, and this makes analyzing the rheology non-trivial. Much like viscoelastic materials, these are characterized by a large diversity of elementary units with multiple characteristic timescales \cite{Tschoegl_book1989,Metzler1996}. There is no universal \emph{predictive} model for thixotropy. Therefore, to enable comparison between materials, we introduce a universal \emph{descriptive} model. Our objective is to develop methods, irrespective of the material tested or the underlying predictive constitutive model, to quantify thixotropy using observed stress changes ($\Delta\sigma$) and timescales ($\tau_{\rm char}$). This concept is illustrated in Fig.~\ref{fig:intro_Ashby}.

% subsection break -----------------------------------------------------

\begin{figure}[!ht]
	\centering
	\includegraphics[scale=0.15,trim={0 0 0 0},clip]{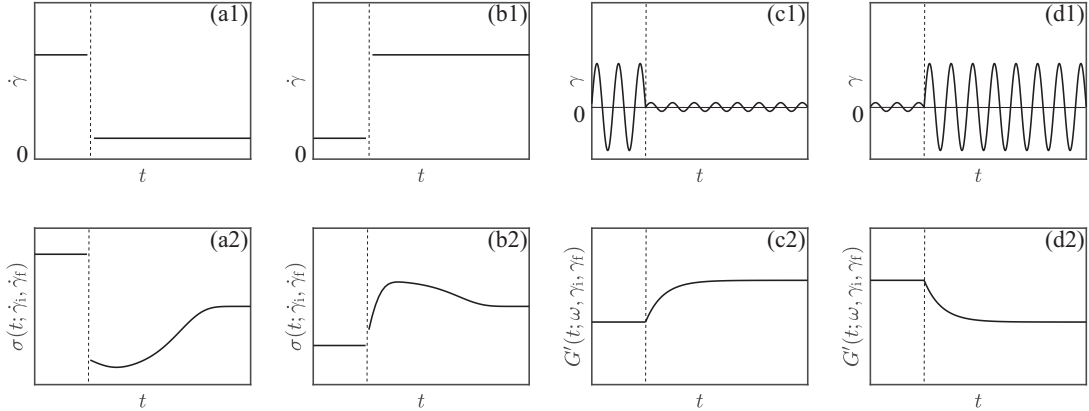}
	\caption{\label{fig:intro_signals}Example input protocols where a thixotropic spectrum of timescales can be applied. (a1) Step down in steady shear ($\dot{\gamma}_{\rm f}<\dot{\gamma}_{\rm i}$), (b1) step up in steady shear ($\dot{\gamma}_{\rm f}>\dot{\gamma}_{\rm i}$), (c1) step down in oscillatory shear amplitude ($\gamma_{\rm f}<\gamma_{\rm i}$), (d1) step up in oscillatory shear amplitude ($\gamma_{\rm f}>\gamma_{\rm i}$). Typical response functions (stresses and moduli) are shown in (a2)-(d2).}
\end{figure}

The simple idea of using $\Delta\sigma$ and $\tau_{\rm char}$ actually involves underlying high-dimensionality, noted in Fig.~\ref{fig:intro_Ashby}. Both thixotropic breakdown and buildup processes must be considered, and the associated kinetics warrant the use of a distribution of timescales to describe the data, eventually reduced to obtain a secondary, average metric of $\tau_{\rm char}$. The amount of thixotropic change depends on the experimental protocol used; the four most common ones are shown in Fig.~\ref{fig:intro_signals}. Using shear stress to study the changes (and hence $\Delta\sigma$) allows us to include information about viscosity (rate-controlled tests) or modulus (oscillatory tests). Lastly, both $\Delta\sigma$ and $\tau_{\rm char}$ are functions of the specific experimental conditions native to the chosen protocol. In step shear rate tests, $\Delta\sigma$ and $\tau_{\rm char}$ are functions of the initial ($\dot{\gamma}_{\rm i}$) and final ($\dot{\gamma}_{\rm f}$) shear rates, or initial ($\gamma_{\rm i}$) and final ($\gamma_{\rm f}$) shear strain amplitudes for oscillatory amplitude step shear (oscillatory) tests. Fig.~\ref{fig:intro_Ashby} is thus a two-dimensional projection of this high-dimensional space onto a plane. On this plane we can still vary $\dot{\gamma}_{\rm i}$ and $\dot{\gamma}_{\rm f}$. Other metrics in addition to $\Delta\sigma$ and $\tau_{\rm char}$ may also exist, as noted in Fig.~\ref{fig:intro_Ashby}.

% subsection break -----------------------------------------------------

Here we introduce the concept of thixotropic spectra, revealed by step changes, as a description of thixotropic effects. Step changes, rather than hysteresis loops, allow thixotropic spectra to be revealed. Many other step tests are available (Fig.~\ref{fig:intro_signals}), but we focus our attention on step shear rate since this can distinguish thixotropy and viscoelasticity \cite{MewisWagner_book2012}. It may be a step down ($\dot{\gamma}_{\rm f} < \dot{\gamma}_{\rm i}$) or step up ($\dot{\gamma}_{\rm f} > \dot{\gamma}_{\rm i}$) test, and the transient stress evolution is used to study thixotropic structure buildup or breakdown. Fig.~\ref{fig:intro_signals}(a) and (b) show typical stress signals for step down (buildup) and step up (breakdown) rate tests respectively. In step down tests, the stress may first show a viscoelastic relaxation (decrease) at very short times, which may or may not be observable \cite{MewisReview1979,MewisWagnerReview2009}. This may be followed by thixotropic recovery at long times \cite{MewisReview1979,MewisWagnerReview2009}. In step up tests, the stress typically shows a viscoelastic stress increase at short times, followed by thixotropic decay at long times \cite{MewisReview1979,MewisWagnerReview2009}. For a given test, the data is analyzed to obtain necessary information about $\Delta\sigma(\dot{\gamma}_{\rm i},\dot{\gamma}_{\rm f})$ and $\tau_{\rm char}(\dot{\gamma}_{\rm i},\dot{\gamma}_{\rm f})$ while keeping in mind the dimensionality of the problem. The analysis method is dictated by the protocol, both of which are described in detail in the following section.

It must be noted that our approach only describes the thixotropic component of the stress response; it requires the viscoelastic component to be sub-dominant or negligible. As needed, the early-time viscoelastic response (as shown schematically in Fig.~\ref{fig:intro_signals}(a2),~(b2)) can be fit to a viscoelastic spectrum superimposed on the thixotropic one. This will truncate the range of observable thixotropic timescales. In our work, we only fit the thixotropic part following any initial viscoelastic response.

%%%%%%%%%%%%%%%%%%%%%%%%%%%%%%%%%%%%%%%%%%%%%%%%%%%%%%%%%%%%%%%%%%%%%%%%%%%%%%%%
%===============================================================================
% section break
%===============================================================================
%%%%%%%%%%%%%%%%%%%%%%%%%%%%%%%%%%%%%%%%%%%%%%%%%%%%%%%%%%%%%%%%%%%%%%%%%%%%%%%%

\section{Theory\label{sec:theory}}

\subsection{Flow conditions and rheometry protocols\label{subsec:theory-protocol}}
We propose describing data in step tests as a superposition of stress components evolving over a spectrum of thixotropic buildup or breakdown timescales as a universal way to quantify thixotropic dynamics. The transient thixotropic stress signal $\sigma$ evolving over time $t$, expressed as a generalized discrete spectrum, is
\begin{subequations}\label{eq:discspec-intro}
	\begin{align}
		\sigma^+(t;\mathbb{P}) &= \sigma_0 + \sum\limits_{i=1}^N \sigma^+_i \left( 1 - e^{-t/\tau^+_i} \right) \label{eq:discspec-intro-recovery},\\
		\sigma^-(t;\mathbb{P}) &= \sigma_{\rm ss} + \sum\limits_{i=1}^N \sigma^-_i e^{-t/\tau^-_i} \label{eq:discspec-intro-breakdown},
	\end{align}
\end{subequations}
where ``$+$'' is used for recovery (step down) and ``$-$'' for breakdown (step up). Subscripts ``0'' and ``ss'' refer to initial $(t=0)$ and steady state $(t\rightarrow\infty)$ values respectively. We assume the basis functions to be \emph{exponential} for each individual mode. It is shown here for $N$ discrete stress modes with mode strengths $\sigma_i$ and associated timescales $\tau_i$. These modes can be either for recovery or breakdown, and superscripts used to distinguish the two as $\tau^+_i$ or $\tau^-_i$, but one would omit the ``$\pm$'' superscripts for generalized discussions. $\mathbb{P}$ is the set of experimental parameters for a given protocol (e.g.\ $\mathbb{P} = \{\dot{\gamma}_{\rm i}$,$\dot{\gamma}_{\rm f}\}$ in step rate tests, a steady state at $\{\dot{\gamma}_{\rm i}$ was achieved). The distribution of $\sigma_i$ over $\tau_i$ can be reduced further to generate low-dimensional summarizing metrics, as discussed in \S~\ref{subsubsec:spectra-moments}. The key idea is to represent a signal as the sum (or integral) of multiple component signals, each with its own characteristic timescale of evolution. It can be applied to step changes of unidirectional or oscillatory shear forcing; the input scheduling for each shown in Fig.~\ref{fig:intro_signals}.

Step rate tests are most commonly used for studying thixotropy \cite{MewisWagnerReview2009}. The exact form of Eq.~\ref{eq:discspec-intro} applied to step rate tests is
\begin{subequations}\label{eq:spectra-steprate}
	\begin{align}
		\sigma^+(t;\dot{\gamma}_{\rm i},\dot{\gamma}_{\rm f}) &= \sigma_0 + \sum\limits_{i=1}^N \sigma^+_i \left( 1 - e^{-t/\tau^+_i} \right) \label{eq:spectra-steprate-down},\\
		\sigma^-(t;\dot{\gamma}_{\rm i},\dot{\gamma}_{\rm f}) &= \sigma_{\rm ss} + \sum\limits_{i=1}^N \sigma^-_i e^{-t/\tau^-_i} \label{eq:spectra-steprate-up},
	\end{align}
\end{subequations}
for recovery and breakdown respectively. We assume that the system has reached steady state during $\dot{\gamma}_{\rm i}$, and time at the step from $\dot{\gamma}_{\rm i}$ to $\dot{\gamma}_{\rm f}$ does not matter; $t = 0$ is measured from the beginning of the $\dot{\gamma}_{\rm f}$ step.

The same idea can be applied to oscillatory shear with step strain amplitude. From the oscillatory signal, a chosen stress amplitude component evolves as
\begin{subequations}
	\begin{align}\label{eq:spectra-stepstrain}
		\sigma^+(t;\omega,\gamma_{\rm i},\gamma_{\rm f}) &= \gamma_{\rm f} G_0(\omega) + \gamma_{\rm f}\sum\limits_{i=1}^N G^+_i(\omega) \left( 1 - e^{-t/\tau^+_i} \right),\\
		\sigma^-(t;\omega,\gamma_{\rm i},\gamma_{\rm f}) &= \gamma_{\rm f} G_{\rm ss}(\omega) + \gamma_{\rm f}\sum\limits_{i=1}^N G^-_i(\omega) e^{-t/\tau^-_i},
	\end{align}
\end{subequations}
for recovery and breakdown respectively. The step in strain amplitude occurs between oscillatory shear inputs $\gamma(t) = \gamma_{\rm i} \sin (\omega t)$ and $\gamma(t) = \gamma_{\rm f} \sin (\omega t)$. Once again, we assume that the oscillatory stress has reached steady state during $\gamma_{\rm i}$. The stress amplitude could be total stress $\sigma$, or in-phase with strain $\sigma^\prime$, or in-phase with strain rate $\sigma^{\prime\prime}$, in which case the parameters based on $G$ would either be $|G^*|$, or $G^\prime$ (as shown in Fig.\ref{fig:intro_signals}), or $G^{\prime\prime}$, respectively. Other oscillatory measures also fit in this framework (e.g.\ orthogonal superposition, where moduli measured in a direction orthogonal to rotational shear, $G^\prime_\perp$ and $G^{\prime\prime}_\perp$, can be used with step rate inputs \cite{Mewis_OSP,Wang_JoR}). Other generalizations of Eq.~\ref{eq:discspec-intro} are possible. For instance, stress in step rate tests can be normalized to define the material function $\eta_i(\tau_i) \equiv \sigma_i(\tau_i)/\dot{\gamma}_{\rm f}$ and reframe Eq.~\ref{eq:discspec-intro} in terms of $\eta_i^+$ and $\eta_i^-$. We use stresses throughout our work here to keep the analysis general and applicable to both strain and strain rate controlled tests. The following section elaborates the mathematical framework for obtaining the spectra.

% subsection break -----------------------------------------------------

\subsection{Thixotropic spectra: discrete and continuous\label{subsec:spectra-spectra}}
To derive the continuous spectrum representation, consider thixotropic recovery as an example, Eq.~\ref{eq:discspec-intro-recovery}. The transient contribution can be written in an integral form, using the properties of the Dirac delta function, as \cite{LucaRHEMAOS_2018}
\begin{align}\label{eq:disc-cont1}
	\sum\limits_{i=1}^N \sigma^+_i \left( 1-e^{-t/\tau^+_i} \right) = \int\limits_{0}^{\infty}\left[ \lim_{N\rightarrow\infty} \sum\limits_{i=1}^{N} \sigma^+_i \cdot \left( 1-e^{-t/\tau^+} \right) \cdot \delta \left( \tau^+-\tau^+_i \right) \right] {\rm d}\tau^+.
\end{align}
Regrouping the terms, we get
\begin{align}\label{eq:disc-cont2}
	\sum\limits_{i=1}^N \sigma^+_i \left(1-e^{-t/\tau^+_i} \right) = \int\limits_{0}^{\infty}\left[ \lim_{N\rightarrow\infty} \sum\limits_{i=1}^{N} \sigma^+_i \cdot \delta \left( \tau^+-\tau^+_i \right) \right] \left( 1-e^{-t/\tau^+} \right) {\rm d}\tau^+.
\end{align}
The grouping in square braces, in the limit of the number of modes $N\rightarrow\infty$, is the continuous spectrum \cite{Tschoegl_book1989,LucaRHEMAOS_2018} $X^+\left(\tau^+\right)$,
\begin{align}\label{eq:line-contspec-recovery}
	X^+\left(\tau^+\right) \equiv \lim_{N\rightarrow\infty} \sum\limits_{i=1}^{N} \sigma^+_i \cdot \delta \left( \tau^+-\tau^+_i \right),
\end{align}
where $\delta(\tau - \tau_i)$ is the shifted Dirac delta function, with SI units of $\rm{s}^{-1}$ \cite{Tschoegl_book1989}. $X^+\left(\tau^+\right)$ is the distribution of thixotropic recovery modes over a domain of recovery timescales $\tau^+$, each mode of spectral strength $\sigma^+_i$ and associated timescale $\tau^+_i$, i.e.\ a small magnitude of \emph{recovered} stress on \emph{recovery} time. By analogy, from Eq.~\ref{eq:discspec-intro-breakdown}, the continuous spectrum of breakdown timescales is 
\begin{align}\label{eq:line-contspec-breakdown}
	X^-\left(\tau^-\right) \equiv \lim_{N\rightarrow\infty} \sum\limits_{i=1}^{N} \sigma^-_i \cdot \delta \left( \tau^- - \tau^-_i \right).
\end{align}

Integration over the entire domain of $\tau$ yields the total stress from all modes. This establishes an equivalence between the discrete and continuous spectra, e.g.\ for recovery
\begin{align}\label{eq:disc-cont3}
	\lim_{N\rightarrow\infty} \sum\limits_{i=1}^N \sigma^+_i \left( 1-e^{-t/\tau^+_i} \right) = \int\limits_{0}^{\infty} X^+\left(\tau^+\right) \cdot \left( 1-e^{-t/\tau^+} \right) {\rm d}\tau^+.
\end{align}
$X^+\left(\tau^+\right)$ is a distribution over timescales, such that an increment in stress $\delta\sigma^+ = X^+\left(\tau^+\right)\delta\tau^+$. $X$ thus has SI units of Pa~s$^{-1}$. It can also be thought of as a spectrum distributed over logarithmically-spaced time increments, such that $\delta\sigma^+ = \Xi^+\left(\tau^+\right)\delta\ln\tau^+$, which gives
\begin{subequations}
	\begin{align}
		X^+\left(\tau^+\right) \delta\tau^+ &= \Xi^+\left(\tau^+\right) \delta\ln\tau^+,\\
		\implies X^+\left(\tau^+\right) &= \frac{\Xi^+\left(\tau^+\right)}{\tau^+}, \label{eq:stress-to-visc-spec}
	\end{align}
\end{subequations}
and $\Xi^+\left(\tau^+\right)$ has units of Pa. It is useful to recast $X^+\left(\tau^+\right)$ as $\Xi^+\left(\tau^+\right)$ for direct comparison between the discrete and continuous spectra plotted on log-scale $\tau^+$. Note that such comparison is only possible when the discrete spectrum $\sigma^+_i$ is log-spaced in $\tau^+_i$ \cite{LucaRHEMAOS_2018}, which we use in this work (see the following section for details).

The equivalence between $\sigma^+_i$ (discrete) and $\Xi^+$ (continuous) spectra, using Eq.~\ref{eq:stress-to-visc-spec} in Eq.~\ref{eq:disc-cont3}, is
\begin{subequations}\label{eq:disc-cont4}
	\begin{align}
		\lim_{N\rightarrow\infty} \sum\limits_{i=1}^N \sigma^+_i \left( 1-e^{-t/\tau^+_i} \right) &= \int\limits_{0}^{\infty} \frac{\Xi^+\left(\tau^+\right)}{\tau^+} \cdot \left( 1-e^{-t/\tau^+} \right) {\rm d}\tau^+ \\
		&= \int\limits_{-\infty}^{\infty} \Xi^+\left(\tau^+\right) \cdot \left( 1-e^{-t/\tau^+} \right) {\rm d}\ln{\tau^+},
	\end{align}
\end{subequations}
noting the change of variables to ${\rm d}\ln\tau^+$ changes the lower limit of integration. Here, we have focused on recovery with basis $(1-e^{-t/\tau^+_i})$. Breakdown is represented in the same manner, just by changing $(1-e^{-t/\tau^+_i})$ to $e^{-t/\tau^-_i}$ (see Table~\ref{ch2:tab:thixo-VE-compare}).

To summarize, we can now generalize the discrete spectrum of Eq.~\ref{eq:discspec-intro} as continuous spectra. For step down in shear rate, from Eq.~\ref{eq:spectra-steprate-down}, the stress recovery is
\begin{align}\label{eq:contspec-down}
	\sigma^+(t;\dot{\gamma}_{\rm i},\dot{\gamma}_{\rm f}) = \sigma_0 + \int\limits_{-\infty}^{\infty} \Xi^+\left(\tau^+;\dot{\gamma}_{\rm i},\dot{\gamma}_{\rm f}\right) \cdot \left( 1 - e^{-t/\tau^+} \right) {\rm d}\ln\tau^+,
\end{align}
and from Eq.~\ref{eq:spectra-steprate-up}, the stress breakdown is
\begin{align}\label{eq:contspec-up}
	\sigma^-(t;\dot{\gamma}_{\rm i},\dot{\gamma}_{\rm f}) = \sigma_{\rm ss} + \int\limits_{-\infty}^{\infty} \Xi^-\left(\tau^-;\dot{\gamma}_{\rm i},\dot{\gamma}_{\rm f}\right) \cdot e^{-t/\tau^-} {\rm d}\ln\tau^-.
\end{align}

% subsection break -----------------------------------------------------

\subsection{Parameterized continuous spectra\label{subsec:theory-contspec}}
The particular shape of the continuous distribution $\Xi(\tau)$ is unspecified in the general theory. Different shapes and parameterized continuous spectra may be possible, similar to viscoelastic spectra, e.g.\ BSW, Lorentzian, fractional Maxwell, Weibull distributions, and their numerous variations, to name a few \cite{LucaRHE_TSS2019}. One may fit either discrete or continuous spectra based on experimental observations (data fitting), or theoretical considerations from microstructure-based models. In the past, stretched exponential forms of stress recovery/decay have been widely used to phenomenologically describe thixotropic transients in step shear tests \cite{DullaertMewis_structkinetics2006,MewisWagnerReview2009,WeiSolomonLarson_SE-JOR2016,WeiSolomonLarson-JOR2018,LarsonWei-JoR2019}. Wei \emph{et al.}, in a series of papers, used a structure kinetics based constitutive model to study thixotropy in aggregating systems, where they employed a stretched exponential distribution of thixotropic structure parameters over a domain of characteristic thixotropic rate constants \cite{WeiSolomonLarson_SE-JOR2016,WeiSolomonLarson-JOR2018}. This idea of a distribution of structure parameters over characteristic rate constants is similar to our approach, except that we directly describe the timescales (rather than assuming a kinetic rate equation) and allow any shape of the distribution (rather than assuming stretched exponential) of stress modes to directly quantify thixotropic observations.

One could use the stretched exponential form to fit recovery/breakdown data, and our framework enables its interpretation as a thixotropic continuous spectrum \cite{Johnston_SE-PRB2006,Santos2005}. Given this common practice, we also show fits to the stretched exponential recovery $(+)$ or decay $(-)$, given by
\begin{subequations}
	\begin{align}\label{eq:strexp_stress}
		\sigma^+(t) &= \sigma_0 + \sigma^+_{\rm se} \left[ 1 - e^{- \left( t/\tau^+_{\rm se} \right)^\beta} \right],\\
		\sigma^-(t) &= \sigma_{\rm ss} + \sigma^-_{\rm se} e^{- \left( t/\tau^-_{\rm se} \right)^\beta},
	\end{align}
\end{subequations}
where $\sigma_{\rm se}$ is the total amount of stress change, happening over a timescale $\tau_{\rm se}$. The stretched exponential function can be related to an underlying continuous distribution of single-exponential modes \cite{Johnston_SE-PRB2006,Santos2005,WeiSolomonLarson_SE-JOR2016,WeiSolomonLarson-JOR2018}. It can therefore be written as a continuous distribution in our framework, given by \cite{Johnston_SE-PRB2006,Santos2005} (see Supplementary Information \S~1 for derivation)
\begin{align}\label{eq:strexp_spectrum}
	\Xi\left(\tau\right) &\equiv \sigma_{\rm se} \frac{1}{\pi} \frac{\tau_{\rm se}}{\tau} \int\limits_{0}^{\infty} e^{-u^{\beta} \cos \left( \pi\beta/2 \right)} \cdot \cos\left[ \frac{\tau_{\rm se}}{\tau}u - u^\beta \sin \left( \frac{\pi\beta}{2} \right) \right] {\rm d}u.
\end{align}
The mathematical derivation takes care to distinguish between $X(\tau)$ and $\Xi(\tau)$. We will use $\Xi$ to compare to discrete thixotropic spectra with log-spaced $\tau_i$. Note that using the stretched exponential $\Xi(\tau)$ to describe step shear data directly is \emph{not} equivalent to the multi-lambda kinetic rate equation used by Wei \emph{et al.}

Among the many possible shapes of $\Xi(\tau)$, we also consider a log-normal distribution. It is given by \cite{LucaRHEMAOS_2018,LucaRHE_TSS2019}
\begin{align}\label{eq:logn}
\Xi(\tau) = \Xi_{\rm m} \exp \left[ - \frac{1}{2} \left(\frac{\ln\tau - \ln\tau_{\rm m}}{\theta} \right)^2 \right],
\end{align}
where $\Xi_{\rm m}$, $\tau_{\rm m}$, and $\theta$ are parameters of the distribution pertaining to the peak value (strength of stress change), log-median relaxation timescale (mean timescale of change), and standard deviation of the spectrum (breadth of the distribution) respectively. This distribution captures the key ideas of thixotropic spectra (a dominant timescale and breadth of distribution) using just three parameters. It will be used to build intuition for comparing discrete and continuous spectra.

% subsection break -----------------------------------------------------

\subsection{Reduced parameters: moments and timescales\label{subsubsec:spectra-moments}}
The moments of the discrete $(\tau_i,\sigma_i)$ and the continuous distribution $\Xi\left(\tau\right)$ are defined as \cite{Tschoegl_book1989,LucaRHEMAOS_2018}
\begin{subequations}\label{eq:moments-define}
\begin{align}
M_n &\equiv \sum_{i=1}^{N} \tau^n \cdot \sigma_i,\\
M_n &\equiv \int\limits_{0}^{\infty} \tau^n \cdot X\left(\tau\right) {\rm d}\tau \equiv  \int\limits_{-\infty}^{\infty} \tau^n \cdot \Xi\left(\tau\right) {\rm d}\ln\tau,
\end{align}
\end{subequations}
respectively, and so on, for $n \in \mathbb{Z}$. Based on these moments, meaningful low-dimensional metrics, such as characteristic timescales of recovery (or breakdown) can be defined. The general definition of the $n$-th mean timescale of a distribution is \cite{Tschoegl_book1989,LucaRHEMAOS_2018}
\begin{align}\label{eq:moments-taun-define}
	\tau_n &\equiv \frac{M_n}{M_{n-1}}.
\end{align}

We propose the use of Ashby-style diagrams \cite{Ashby_book2011} (\S~\ref{sec:lowdim}) for plotting the quantities defined here. One meaningful quantity is $M_0$, the net change in stress during the recovery/breakdown process. We denote this total change in stress as $\Delta\sigma$,
\begin{subequations}\label{eq:moments-M0-deltasigma}
\begin{align}
	\Delta\sigma &= M_0 \equiv \sum_{i=1}^{N} \sigma_i,\\
	\Delta\sigma &= M_0 \equiv \int\limits_{-\infty}^{\infty} \Xi\left(\tau\right) {\rm d}\ln\tau,
\end{align}
\end{subequations}
respectively, for the discrete and continuous distributions. The average timescales $\tau_n$ are also meaningful. In terms of the spectra, Eq.~\ref{eq:moments-taun-define} is
\begin{subequations}\label{eq:taun}
\begin{align}
	\tau_n = \frac{M_n}{M_{n-1}} &\equiv \dfrac{\sum\limits_{i=1}^N \tau_i^n \cdot \sigma_i}{\sum\limits_{i=1}^N \tau_i^{n-1} \cdot \sigma_i},\\
	\tau_n = \frac{M_n}{M_{n-1}} &\equiv \dfrac{\int\limits_{-\infty}^{\infty} \tau^n \cdot \Xi\left(\tau\right) {\rm d}\ln\tau}{\int\limits_{-\infty}^{\infty} \tau^{n-1} \cdot \Xi\left(\tau\right) {\rm d}\ln\tau},
\end{align}
\end{subequations}
respectively, for the discrete and continuous distributions. From this, we can define three important quantities: the first and second mean timescales, and a polydispersity of the timescales, as
\begin{subequations}
	\label{eq:moments-tau-define}
	\begin{align}
		\tau_1 &\equiv \frac{M_1}{M_0}, \label{eq:moments-tau-define-tau1}\\
		\tau_2 &\equiv \frac{M_2}{M_1}, \label{eq:moments-tau-define-tau2}\\
		\rm{PDI} &\equiv \frac{\tau_2}{\tau_1} = \frac{M_2M_0}{M_1^2}.  \label{eq:moments-tau-define-PDI}
	\end{align}
\end{subequations}
There are other ways to define mean timescales. For e.g., one may define mean timescales in the log space, such as
\begin{align}\label{eq:moments-tau-define-log}
	\ln\tau_{n,{\rm log}} &\equiv \dfrac{\int\limits_{-\infty}^{\infty} (\ln\tau)^n \cdot \Xi\left(\tau\right) {\rm d}\ln\tau}{\int\limits_{-\infty}^{\infty} (\ln\tau)^{n-1} \cdot \Xi\left(\tau\right) {\rm d}\ln\tau}.
\end{align}
In essence, the definition of mean timescales is a matter of choice. We have defined the mean timescales in Eq.~\ref{eq:moments-tau-define} to mirror the definitions used in linear viscoelastic spectra \cite{Tschoegl_book1989,LucaRHEMAOS_2018}, and shall be using these to populate Ashby charts for thixotropy data. Defining the mean timescales in this manner is akin to a weighted arithmetic average of timescales, while Eq.\ref{eq:moments-tau-define-log} is meaningful for certain distributions such as the log-normal (Eq.~\ref{eq:logn}), where mean timescales defined on log space are equal to log-mean timescales of the specific distribution, $\ln\tau_{n,{\rm log}} = \ln\tau_{\rm m}$.

Eq.~\ref{eq:spectra-steprate-down}, \ref{eq:spectra-steprate-up} (discrete spectra) and \ref{eq:contspec-down}, \ref{eq:contspec-up} (continuous spectra), in conjunction with Eq.~\ref{eq:moments-M0-deltasigma}~and~\ref{eq:taun} to obtain the reduced metrics, convey the core idea of this work. There are similarities between spectra used in viscoelasticity and the thixotropic spectra used here. In linear viscoelasticity, the stress relaxation shear modulus can be expressed in discrete or continuous forms as $G(t) = \sum\limits_{i = 1}^N G_i \cdot e^{-t/\tau_i}$, or $G(t) = \int\limits_0^\infty Q(\tau) \cdot e^{-t/\tau} {\rm d}\tau = \int\limits_{-\infty}^\infty H(\tau) \cdot e^{-t/\tau} {\rm d}\ln\tau$, where $Q(\tau)$ and $H(\tau)$ are the modulus-weighted and viscosity-weighted viscoelastic relaxation spectra, respectively \cite{LucaRHEMAOS_2018}. A summary of thixotropic spectra and the relevant basis functions, with parallels to viscoelastic spectra, is given in Table~\ref{ch2:tab:thixo-VE-compare}.

\begin{table}[!ht]
	\caption{Thixotropic spectra with parallels to viscoelastic spectra. Definitions of $Q(\tau)$ and $H(\tau)$ following Martinetti \emph{et al.} \cite{Tschoegl_book1989,LucaRHEMAOS_2018,LucaRHE_TSS2019}}
	\label{ch2:tab:thixo-VE-compare}
	\begin{tabular}{l|lllll}
	\hline
		Property & Data & Basis function & Interpretation & Discrete & Continuous (on ${\rm d}\tau$)\\
		\hline
		Thixo, $\dot{\gamma}_{\rm f}>\dot{\gamma}_{\rm i}$& $\sigma^-(t;\dot{\gamma}_{\rm i},\dot{\gamma}_{\rm f})$ & $\sigma^-_i e^{-t/\tau^-_i}$ & Breakdown & $\sigma^-_i(\tau^-_i)$ & $X^-(\tau^-)$ \\
		Thixo, $\dot{\gamma}_{\rm f}<\dot{\gamma}_{\rm i}$& $\sigma^+(t;\dot{\gamma}_{\rm i},\dot{\gamma}_{\rm f})$ & $\sigma^+_i \left(1-e^{-t/\tau^+_i}\right)$ & Recovery & $\sigma^+_i(\tau^+_i)$ & $X  ^+(\tau^+)$ \\
		\hline
		Viscoelastic\cite{DPL_vol1} & $G(t;\gamma_0)$ & $G_i e^{-t/\tau_i}$ & Decay & $G_i(\tau_i)$ & $Q(\tau)$\\
		(relaxation) & $\eta^+(t;\dot{\gamma}_0)$ & $\eta_i \left(1-e^{-t/\tau_i}\right)$ & Growth & $\eta_i(\tau_i) = \tau_i G_i(\tau_i)$ & $H(\tau) = \tau Q(\tau)$\\
		\hline
	\end{tabular}
\end{table}

The reduced parameters from using thixotropic spectra, $\Delta\sigma$ and $\tau_n$, obtained from Eq.~\ref{eq:moments-taun-define},~\ref{eq:moments-M0-deltasigma}, can be used to describe any thixotropic data or constitutive model response. Using $\Delta\sigma$ and $\tau_n$, we can now populate Fig.~\ref{fig:intro_Ashby} to report and compare the degree of thixotropy in different materials. Such results are presented in \S~\ref{sec:lowdim}. The application of this framework to step shear data is illustrated through a theoretical example in the following section.

% subsection break -----------------------------------------------------

\subsection{Illustration: relation between continuous, discrete, and reduced metrics\label{subsec:spectra-application}}

\begin{figure}[!ht]
	\begin{minipage}[!ht]{0.49\textwidth}
		(a)
		\centering
		\includegraphics[scale=0.39,trim={0 0 0 0},clip]{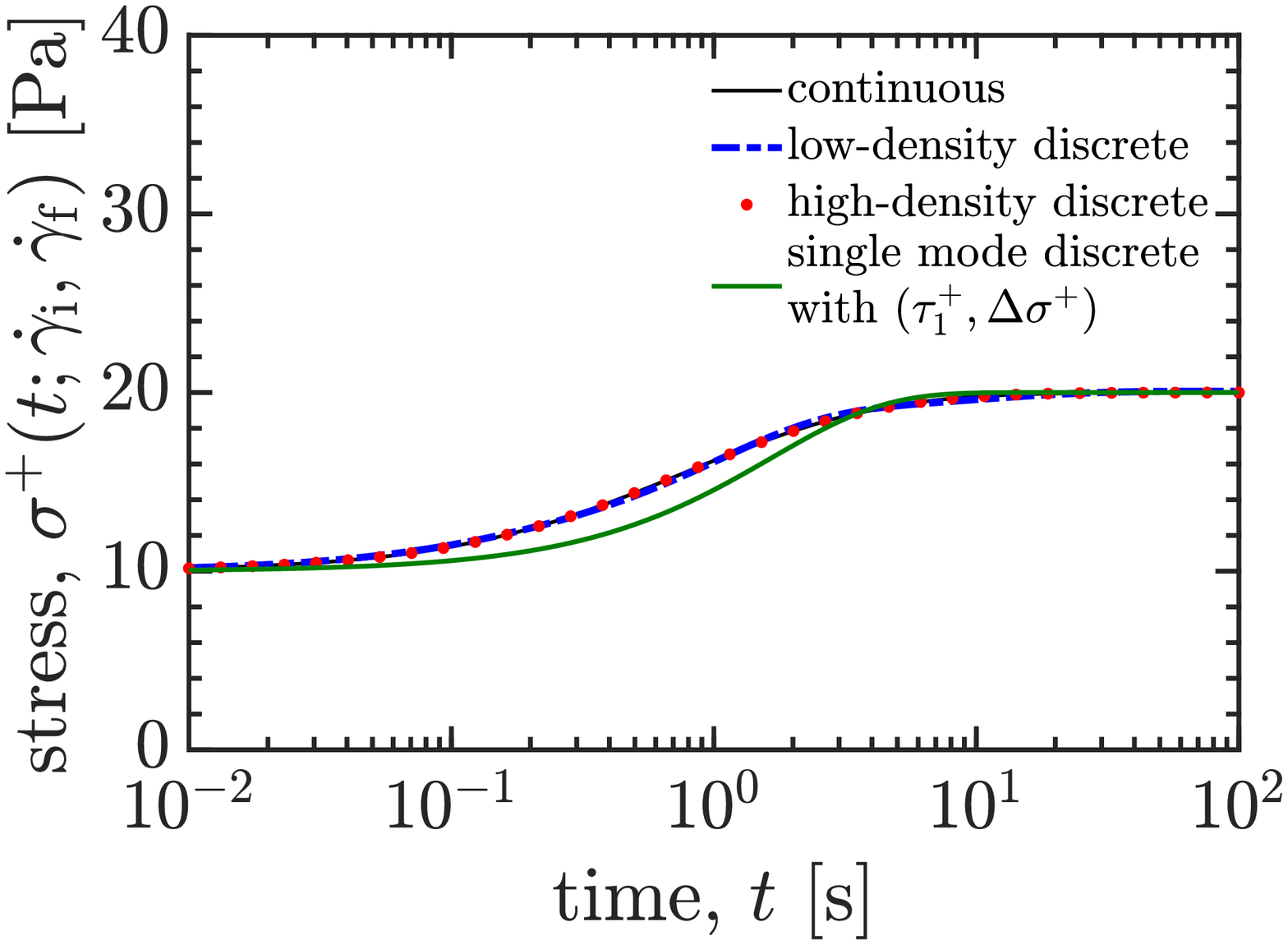}
	\end{minipage}
	\begin{minipage}[!ht]{0.49\textwidth}
		(b)
		\centering
		\includegraphics[scale=0.39,trim={0 0 0 0},clip]{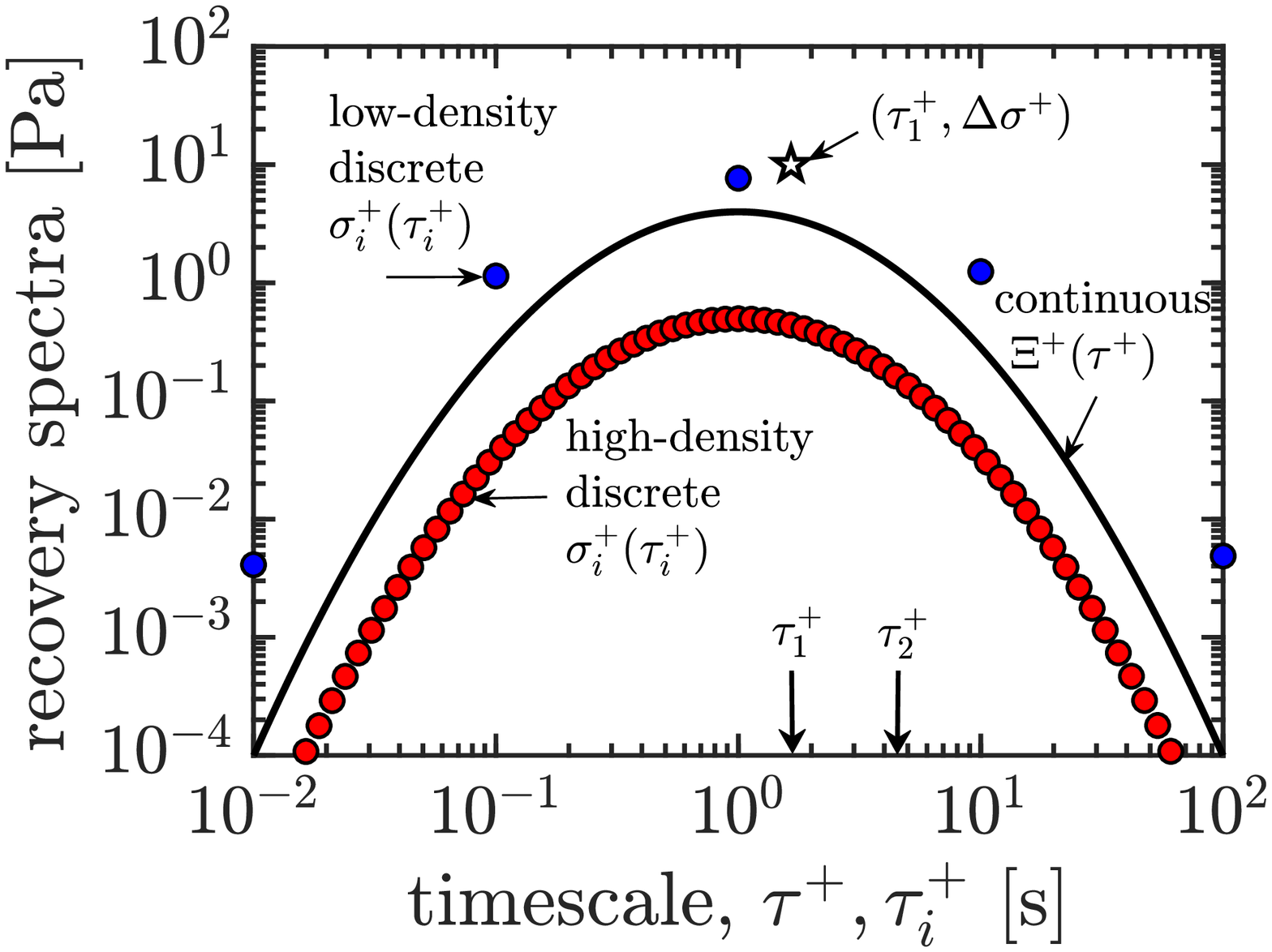}
	\end{minipage}
	\caption{\label{fig:stress-spec_demo}Log-normal thixotropic recovery spectrum compared to discrete spectrum approximations. Simulated data with $\sigma_0 = 10$~Pa, $\Xi_{\rm m} = 10$~Pa, $\tau_{\rm m} = 1$~s, $\theta = 1$, in Eq.~\ref{eq:logn}. (a) Stress response to step down in shear rate, (b) co-plot of each spectrum with important reduced metrics shown.}
\end{figure}

Using the log-normal $\Xi^+(\tau^+)$ of Eq.~\ref{eq:logn} with parameters $\Xi_{\rm m} = 10$~Pa, $\tau_{\rm m} = 1$~s, $\theta = 1$, and $\sigma_0 = 10$~Pa, in Eq.~\ref{eq:contspec-down}, we can produce the typical stress recovery shape in Fig.~\ref{fig:stress-spec_demo}(a). This is similar to the schematic in Fig.~\ref{fig:intro_signals}(a2). The shape of $\Xi^+(\tau^+)$ is shown as the solid line in Fig.~\ref{fig:stress-spec_demo}(b). The log-median timescale $\tau_{\rm m} = 1$~s is associated with the transients shown in Fig.~\ref{fig:stress-spec_demo}(a), while the peak $\Xi_{\rm m} = 10$~Pa is related to the amount of stress change. The PDI of the timescales is related to $\theta$, the spread of the log-normal spectrum \cite{LucaRHEMAOS_2018}. From $\Xi^+$, we obtain the generalized low-dimensional metrics, namely the moments and characteristic timescales, using Eq.~\ref{eq:moments-M0-deltasigma} and \ref{eq:taun}. The timescales $\tau^+_1$ and $\tau^+_2$ for the spectra are shown on the plot. Note that $\tau^+_2 > \tau^+_1$. Also note that $\tau^+_1 = 1.6~{\rm s} > \tau_{\rm m} = 1~{\rm s}$. This is because we calculate mean timescales in $\tau$ space, and not $\ln\tau$ space. If we were to use Eq.~\ref{eq:moments-tau-define-log}, we would get $\tau^+_{1,{\rm log}} = \tau_{\rm m}$. But, as mentioned earlier, this is a matter of choice, and we use mean timescales defined in $\tau$ space similar to those in viscoelasticity.

We also show high-density discrete, and low-density discrete distributions, each co-plotted in Fig.~\ref{fig:stress-spec_demo}(b). The high density (50 modes) and low density (5 modes) distributions $(\sigma^+_i)$ are discrete representations of $\Xi^+$ (obtained by fitting $\sigma^+_i$ over a set domain of $\tau^+_i$ to match the $\sigma^+(t)$ in Fig.~\ref{fig:stress-spec_demo}(a) resulting from the continuous distribution). Also note that $\sigma^+_i$ lies below $\Xi^+$ for a sufficiently large number of modes, whereas for fewer modes the values of $\sigma^+_i$ lie above $\Xi^+$. This is a consequence of the fact that each distribution must result in the same stress signal when integrated (or summed). As more modes are added, the value of each mode must decrease in order to add up to the same stress response. The plot also shows a single data point $(\tau^+_1,\Delta\sigma^+)$, which is the pair of reduced or summarizing metrics obtained using Eq.~\ref{eq:moments-M0-deltasigma} and \ref{eq:taun}, in essence is a single-mode representation of the polydisperse system. This can be plotted in the thixotropic Ashby charts. It lies above all other distributions in Fig.~\ref{fig:stress-spec_demo}(b) as expected. This can be visualized as a single-mode stress response $\sigma^+(t) = \sigma_0 + \Delta\sigma^+\left(1-e^{-t/\tau^+_1}\right)$, shown in Fig.~\ref{fig:stress-spec_demo}(a). This clearly shows that $\tau^+_1 \neq \tau_{\rm m}$, and $\tau^+_1$ is indeed a mean timescale of the original stress response since the $(\tau^+_1,\Delta\sigma^+)$ mode goes through the more complex $\sigma^+(t)$ response. It must be noted that the stress generated from $(\tau^+_1,\Delta\sigma^+)$ is not equivalent to a single-mode discrete spectrum fit of $\sigma^+(t)$ data.

The illustrative example in Fig.~\ref{fig:stress-spec_demo} provides intuition for the breadth of a thixotropic spectrum. Here, ${\rm PDI} = \tau^+_2/\tau^+_1 = 2.08$, quantifying the deviation from a single timescale. It shows that discrete spectra can provide excellent descriptions of $\sigma^+(t)$, and that the log-spaced shape of $\sigma_i(\tau_i)$ will be similar to the shape of any underlying $\Xi(\tau)$ representation (if $\tau_i$ are equally spaced in $\ln\tau_i$). In what follows with experimental data, we use discrete spectra, log-spaced in $\tau_i$, to fit $\sigma^+(t)$ data and obtain reduced metrics on a number of thixotropic materials. The materials used and rheometry and fitting methods employed are shown in the next section, and the fitting results are shown in \S~\ref{sec:results-fits-disc}.

%%%%%%%%%%%%%%%%%%%%%%%%%%%%%%%%%%%%%%%%%%%%%%%%%%%%%%%%%%%%%%%%%%%%%%%%%%%%%%%%
%===============================================================================
% section break
%===============================================================================
%%%%%%%%%%%%%%%%%%%%%%%%%%%%%%%%%%%%%%%%%%%%%%%%%%%%%%%%%%%%%%%%%%%%%%%%%%%%%%%%

\section{Experimental: materials and methods\label{sec:matmeth}}

% subsection break -----------------------------------------------------

\subsection{Materials\label{subsec:materials}}
We test two common yield-stress fluids: 4.0~wt\% suspension of Laponite RD colloidal clay particles in water, which forms a sparse physical gel when dispersed, typically identified as a thixotropic material \cite{MewisWagnerReview2009}; and 1.0~wt\% suspension of Carbopol 940 microgel particles in water, a model yield-stress fluid formed of jammed, swollen, crosslinked polyacrylic acid microgel particles, known to exhibit negligible thixotropy \cite{Piau_cpol2007,Divoux_PRL2013}. These suspensions were prepared using methods from the literature \cite{BCB_PhysFluids2015,SSRHE_JFM2020}. We also use step shear data for two model thixotropic yield-stress fluids from the open literature: 3.23~vol\% carbon black suspensions in naftenic oil \cite{DullaertMewis_structkinetics2006}, and 2.9~vol\% fumed silica suspension in paraffin oil with PIB \cite{ArmWag_JOR2016}. Both form thixotropic colloidal dispersions when suspended in their respective media. We show reduced metrics for all the above materials in the Ashby diagrams, but only show detailed step shear rate data for Laponite in the next section for the sake of brevity.

% subsection break -----------------------------------------------------

\subsection{Rheometry methods\label{subsec:methods}}
The data for all materials is either a step down from an initial shear rate of $\dot{\gamma}_{\rm i} = 5.00~\rm{s}^{-1}$ to various lower final shear rates $\dot{\gamma}_{\rm f} = 0.25,~0.50,~1.00,~2.50~{\rm s}^{-1}$, or a step up from $\dot{\gamma}_{\rm i} = 0.10~\rm{s}^{-1}$ to various higher shear rates $\dot{\gamma}_{\rm f} = 0.50,~1.00,~2.50,~5.00~{\rm s}^{-1}$ (as shown in Fig.~\ref{fig:intro_signals}(a)~and~(b) respectively). Data for fumed silica and carbon black were used directly from the published literature. For Laponite and Carbopol, the materials were sheared between 600 grit sandpaper-covered parallel plate geometries (uncorrected for parallel plate shear rate inhomogeneity), 25~mm in diameter, at two different gaps (750 and 1000~$\mu{\rm m}$) to verify the absence of wall slip \cite{RHE_baddata}, at $25^\circ$C, on an ARES-G2 strain-controlled, separated motor-transducer rheometer (TA Instruments).

% subsection break -----------------------------------------------------

\subsection{Fitting methods\label{subsec:fitting}}
The model fits were done using iterative linear regression algorithms that in their most general form minimize a weighted residual sum of squares, ${\rm RSS} = \sum\limits_{i=1}^d \left[ y_i - f(x_i) \right]^2/w_i$, where $y_i$ are the experimental data (shear stress $\sigma(t)$) and $f(x_i)$ are the model predictions over the control variable (time $t$). The residuals were weighted by the experimental data, so that $w_i = y_i$, which implicitly assumes an error proportional to the data \cite{Singh_2019}. For the stretched exponential fits, the data is fit to Eq.~\ref{eq:strexp_stress}, and Eq.~\ref{eq:strexp_spectrum} is used to interpret the fit as a continuous spectrum. For the discrete spectrum fits, obtaining the underlying distribution is an ill-posed, inverse problem, and we employed Tikhonov regularization \cite{Tikhonov_1979,Kont_2010,Kontogiorgos,WeiSolomonLarson-JOR2018} to obtain the optimal set of $\sigma_i$ for a pre-distributed, log-spaced set of $\tau_i$ values (using open-source MATLAB scripts available online \cite{PCH}).

%%%%%%%%%%%%%%%%%%%%%%%%%%%%%%%%%%%%%%%%%%%%%%%%%%%%%%%%%%%%%%%%%%%%%%%%%%%%%%%%
%===============================================================================
% section break
%===============================================================================
%%%%%%%%%%%%%%%%%%%%%%%%%%%%%%%%%%%%%%%%%%%%%%%%%%%%%%%%%%%%%%%%%%%%%%%%%%%%%%%%

\section{Spectra from experimental data\label{sec:results-fits-disc}}
We first show results for step up tests, followed by step down tests. We co-plot the fits to the discrete spectra and the stretched exponential for the transient stress data, and hence also co-plot the discrete and continuous spectra (corresponding to the underlying spectra for the stretched exponential). The fit parameters for the stretched exponential for each dataset ($\sigma_{\rm se}$ and $\tau_{\rm se}$) are shown in Supplementary Information, \S~2.

\subsection{Step up in shear rate\label{subsec:results-fits-disc-up}}
\begin{figure}[!ht]
	\begin{minipage}[!ht]{0.49\textwidth}
		(a)
		\centering
		\includegraphics[scale=0.39,trim={0 0 0 0},clip]{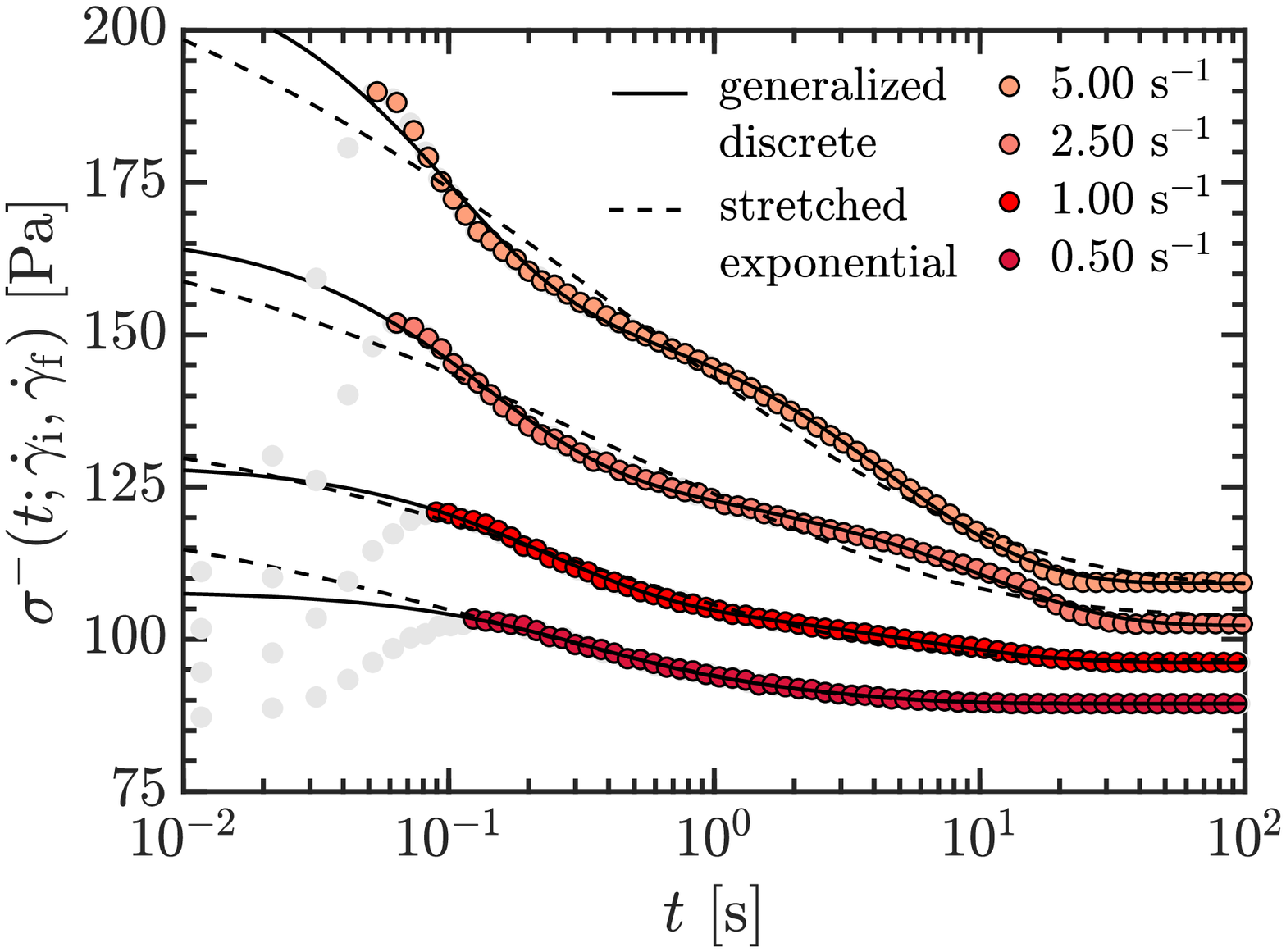}
	\end{minipage}
	\begin{minipage}[!ht]{0.49\textwidth}
		(b)
		\centering
		\includegraphics[scale=0.39,trim={0 0 0 0},clip]{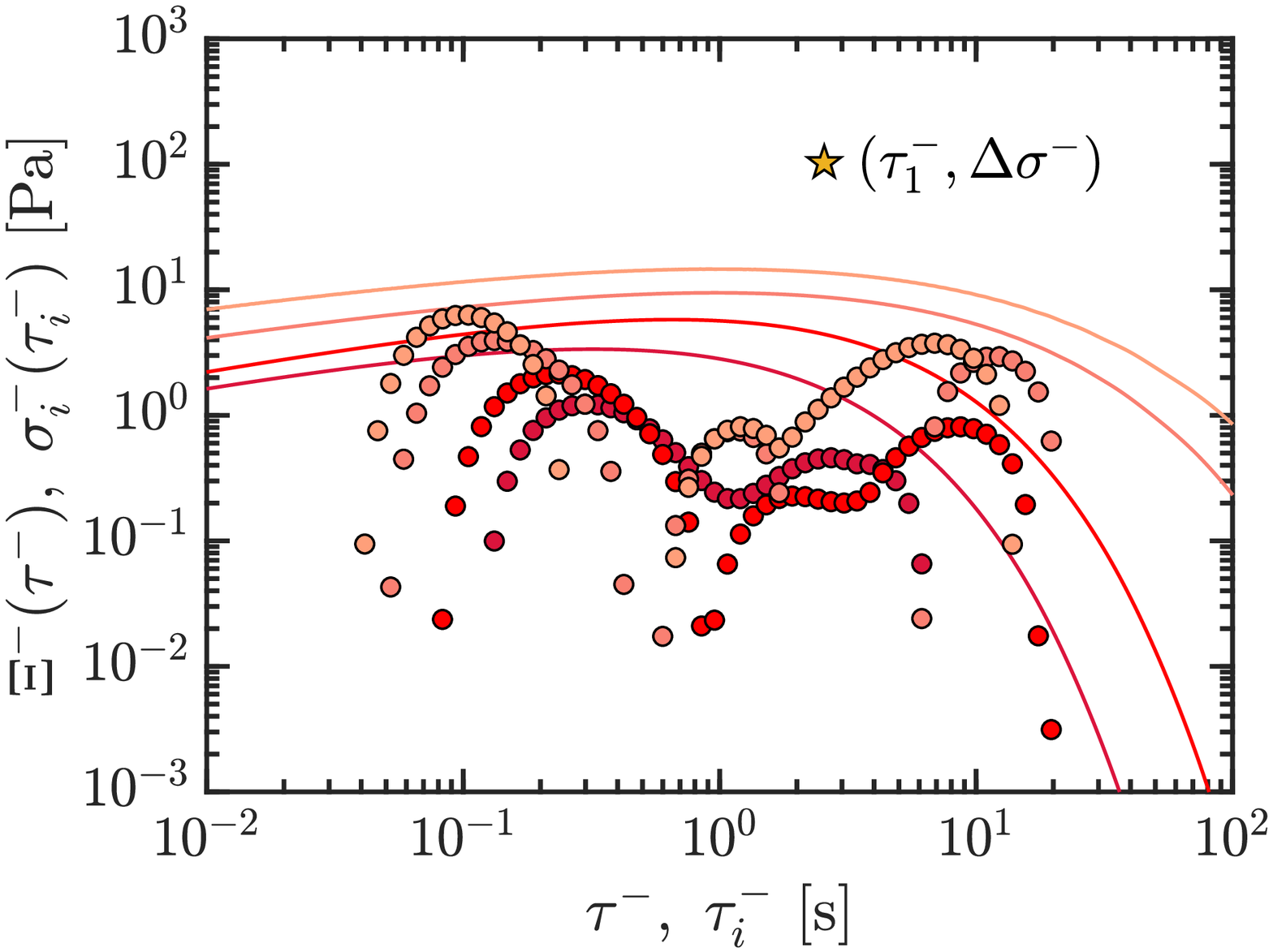}
	\end{minipage}
	\caption{\label{fig:disc_up_stress}Step up in shear rate data for 4.0~wt\% Laponite suspension in water, with $\dot{\gamma}_{\rm i}=0.1~\rm{s}^{-1}$. (a) Transient stress decay fit to a discrete thixotropic spectrum (solid lines) and to a stretched exponential (dashed lines) form of breakdown. (b) Co-plots of the discrete spectrum (circles) and the continuous spectrum interpretation of the stretched exponential (solid lines) for each value of $\dot{\gamma}_{\rm f}$. The $\bigstar$ shows the summarizing metric for the discrete spectrum at the most extreme case of thixotropy, corresponding to $\dot{\gamma}_{\rm f}=5.00~\rm{s}^{-1}$.}
\end{figure}

Fig.~\ref{fig:disc_up_stress} shows step up in shear rate data for 4.0~wt\% Laponite suspension in water, where (a) shows the transient stress data, and fits to the discrete spectrum (Eq.~\ref{eq:spectra-steprate-up}, solid lines) and the stretched exponential (Eq.~\ref{eq:strexp_stress}, dashed lines). The data in (a) suggest that the timescales of breakdown are $\mathcal{O}(10)~{\rm s}$, and the amount of stress change increases from $\approx 20$ to $100~{\rm Pa}$ as $\dot{\gamma}_{\rm f}$ increases. The effect of thixotropy is most extreme for $\dot{\gamma}_{\rm f} = 5.00~{\rm s}^{-1}$.

Thixotropic spectra help quantify these observations. The existence of multiple modes of breakdown, evidenced by the changing concavity in the $\sigma^-(t)$ data, is captured well by the discrete spectrum. The multi-modal nature of the thixotropic breakdown is exemplified in Fig.~\ref{fig:disc_up_stress}(b), where the discrete spectrum $\sigma^-_i(\tau^-_i)$ shows multiple peaks at distinct timescales, $\tau^-_i$. This feature is missing from $\Xi^-\left(\tau^-\right)$ for the single-mode stretched exponential. The summarizing metrics of the discrete spectra follow a systematic trend; both timescales and amount of stress change increase as $\dot{\gamma}_{\rm f}$ increases from 0.50 to 5.00~s$^{-1}$ ($\tau^-_1 = 1.06,~2.23,~2.56,~3.91$~s, and $\Delta\sigma^- = 18.52,~32.64,~64.77,~102.78$~Pa). The degree of thixotropy therefore is the most extreme for the highest $\dot{\gamma}_{\rm f}$, and the corresponding values of $(\tau^-_1,\Delta\sigma^-)$ are co-plotted with the spectra in Fig.~\ref{fig:disc_up_stress}(b). Obtaining such simple yet meaningful descriptions is the key result of this work.

Similar features are retained for step down in shear data as shown in the next section, although the evidence is not conclusive if one only looks at the stress data; one must also look at the underlying spectrum to truly visualize the distribution of characteristic timescales during transient thixotropic processes.

% subsection break -----------------------------------------------------

\subsection{Step down in shear rate\label{subsec:results-fits-disc-down}}

\begin{figure}[!ht]
	\begin{minipage}[!ht]{0.49\textwidth}
		(a)
		\centering
		\includegraphics[scale=0.39,trim={0 0 0 0},clip]{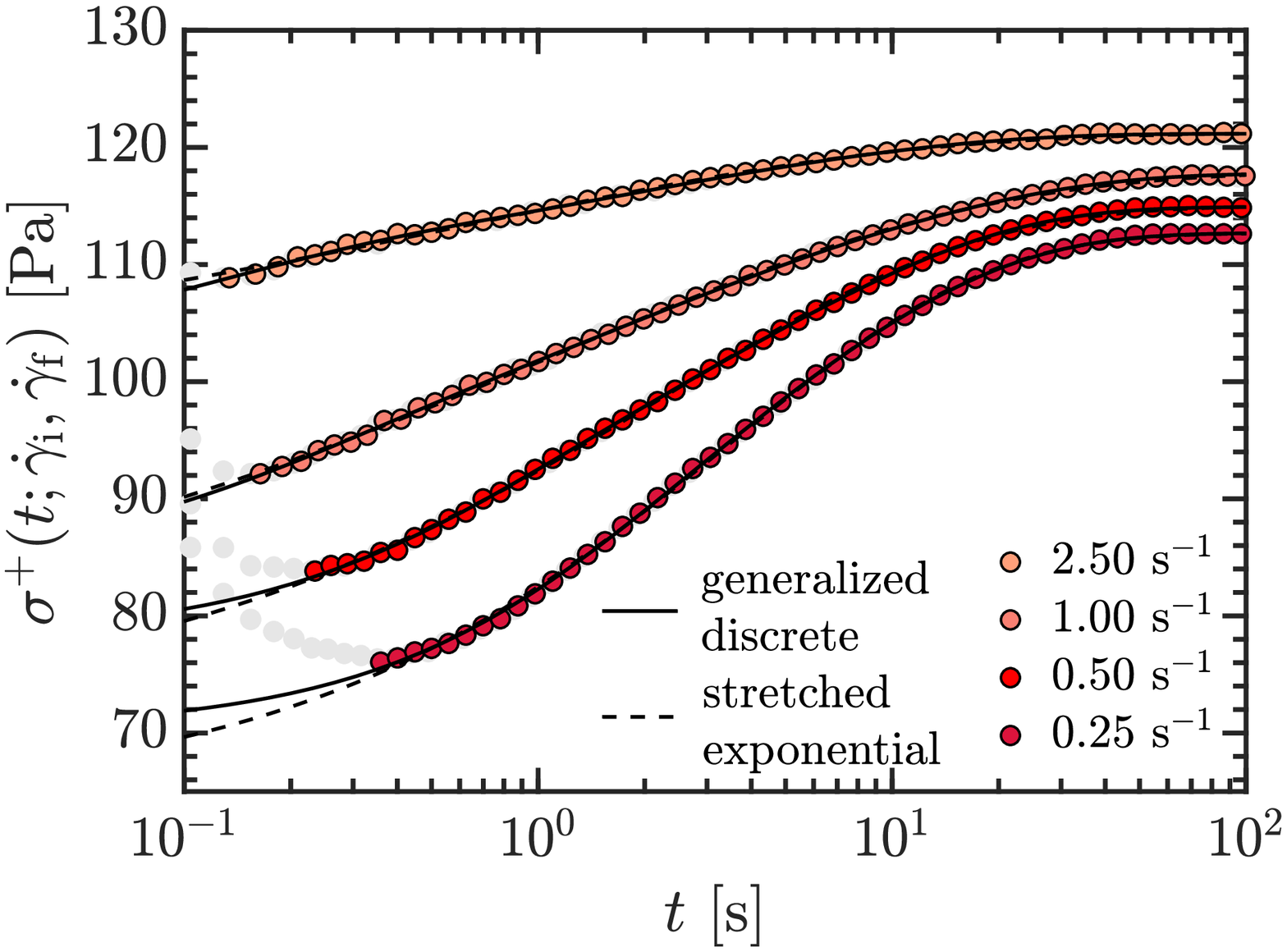}
	\end{minipage}
	\begin{minipage}[!ht]{0.49\textwidth}
		(b)
		\centering
		\includegraphics[scale=0.39,trim={0 0 0 0},clip]{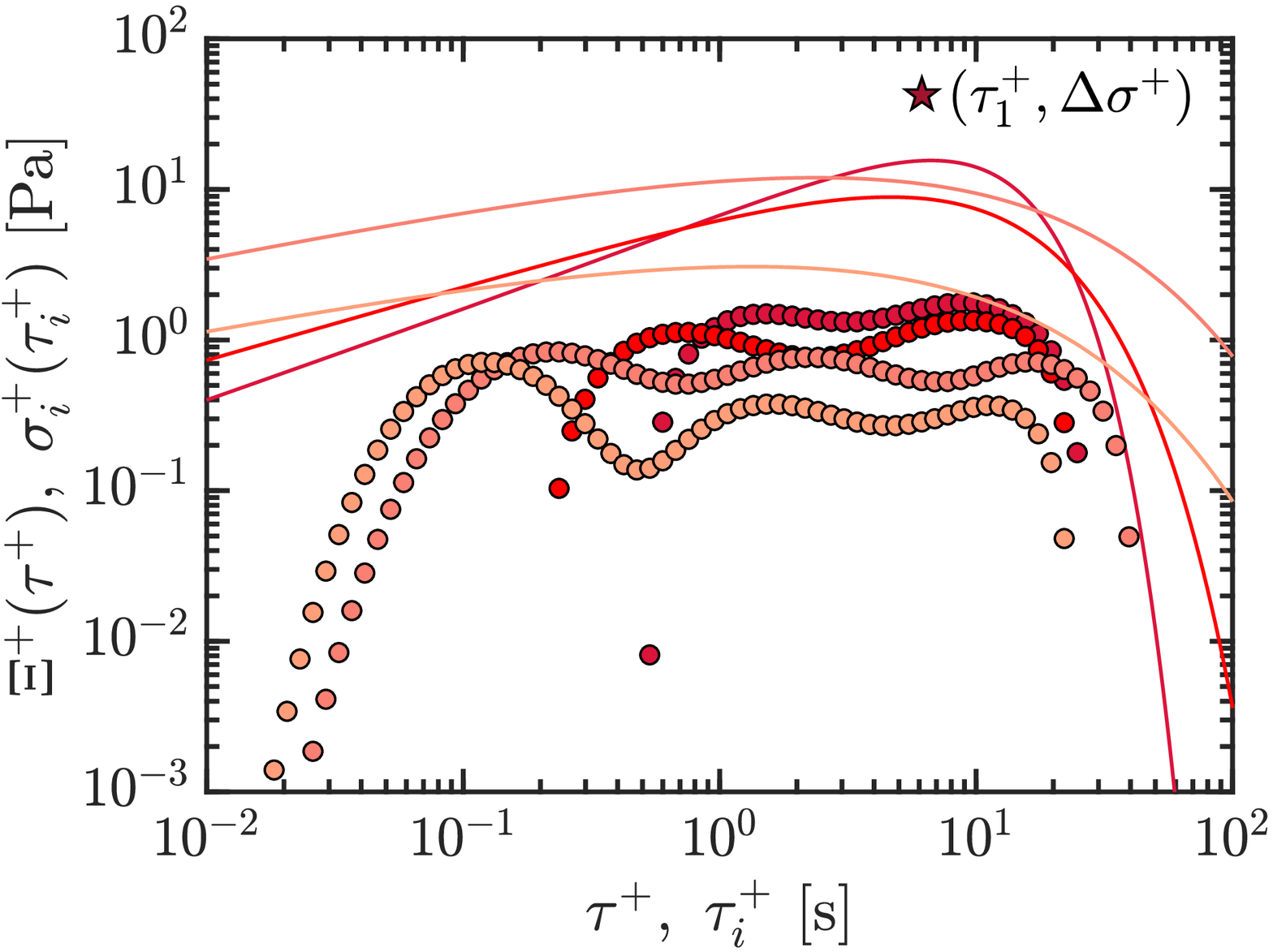}
	\end{minipage}
	\caption{\label{fig:disc_down_stress}Step down in shear rate data for 4.0~wt\% Laponite suspension in water, with $\dot{\gamma}_{\rm i}=5.0~\rm{s}^{-1}$. (a) Transient stress growth fit to a discrete thixotropic spectrum (solid lines) and to a stretched exponential (dashed lines) form of buildup. (b) Co-plots of the discrete spectrum (circles) and the continuous spectrum interpretation of the stretched exponential (solid lines) for each value of $\dot{\gamma}_{\rm f}$. The $\bigstar$ shows the summarizing metric for the discrete spectrum at the most extreme case of thixotropy, corresponding to $\dot{\gamma}_{\rm f}=0.25~\rm{s}^{-1}$.}
\end{figure}

Fig.~\ref{fig:disc_down_stress} shows step down in shear data for 4.0~wt\% Laponite suspension in water, where (a) shows the transient stress data, and fits to the discrete spectrum (Eq.~\ref{eq:spectra-steprate-down}, solid lines) and the stretched exponential (Eq.~\ref{eq:strexp_stress}, dashed lines). The difference between the discrete \emph{versus} continuous distributions is not apparent from the stress data; only from (b) do we see the existence of multiple modes of recovery, which a single-timescale stretched exponential function cannot capture. As for the step up data, the multi-modal nature of the thixotropic recovery is exemplified in Fig.~\ref{fig:disc_down_stress}(b), which is captured by $\sigma^+_i(\tau^+_i)$ but missed by $\Xi^+\left(\tau^+\right)$.

The effect of $\dot{\gamma}_{\rm f}$ is more prominent for the recovery spectra, where the peaks in $\sigma^+_i$ shift toward longer timescales, indicating that the stress recovery takes longer at lower shear rates. This trend is also observed in the stretched exponential $\Xi^+$. This is reflected in the summarizing metrics, which again follow a systematic trend; both timescale and amount of stress recovered increase as $\dot{\gamma}_{\rm f}$ decreases from 2.50 to 0.25~s$^{-1}$ ($\tau_1^+ = 2.84,~5.20,~5.21,~6.14$~s, and $\Delta\sigma^+ = 18.99,~33.63,~36.83,~42.38$~Pa). The degree of thixotropy is the most extreme for the lowest $\dot{\gamma}_{\rm f}$, and the corresponding values of $(\tau_1^+,\Delta\sigma^+)$ are co-plotted with the spectra in Fig.~\ref{fig:disc_down_stress}(b). Such a monotonic trend of increasing recovery timescales with decrease in shear rate is also predicted by classical thixotropic constitutive equations \cite{Goodeve1938,Moore1959,MewisReview1979,LarsonWei-JoR2019} (see Supplementary Information, ~\S~3 for an example), and it is indeed a satisfying if not surprising result that the same trend exists in this dataset.

Note that during any thixotropic process, breakdown or recovery, both shear-induced hydrodynamic forces/stresses (which tend to break up particle aggregates) and Brownian motion due to thermal fluctuations (which helps particles find minima in the interparticle potential energy landscape and assist floc growth) contribute to the net resultant transient stress evolution, i.e.\ the $\sigma(t)$ data shown in the plots. The relative strength of these two effects is often compared using the P\'eclet number \cite{MewisWagner_book2012}, defined as
\begin{align}\label{eq:Peclet}
	{\rm Pe} \equiv \frac{6\pi\eta_{\rm m}\dot{\gamma}_{\rm c}a^3}{k_{\rm B}T} \sim \frac{\text{advective transport rate}}{\text{diffusive transport rate}},
\end{align}
where $\eta_{\rm m}$ is the viscosity of the suspending medium, $a$ is the primary particle size, and $T$ is the absolute temperature of the system. For our step rate tests, the characteristic rate of shear in the system, $\dot{\gamma}_{\rm c} = \dot{\gamma}_{\rm f}$. The strength of Brownian hydrodynamics is independent of shear rate. At high $\dot{\gamma}_{\rm f}$, shear-induced stresses dominate, and the effect of intrinsic particle interactions and Brownian motion is limited. This is the case during step up tests, and advective hydrodynamic stresses dictate the rate of structure breakdown. On the other hand, in step down tests, the thixotropic structure buildup is primarily influenced by interparticle forces and Brownian motion, and less so by advective hydrodynamics, at very low values of $\dot{\gamma}_{\rm f}$ such that $\rm Pe \ll 1$. The resultant recovery processes take much longer and the thixotropic transients are more easily observable. We therefore choose to compare the thixotropic properties of Laponite to other fluids by looking at recovery in step down in shear data in \S~\ref{sec:lowdim}. We show step down in shear data for other materials in the next subsection.

Also note that this limit of $\rm Pe \ll 1$ for $\dot{\gamma}_{\rm f} \rightarrow 0$ is a key feature to report, because it represents the most extreme limit of thixotropic recovery. Using the simplest, classical thixotropic constitutive equations \cite{Goodeve1938,Moore1959,MewisReview1979}, each timescale mode $\tau^+_i$ can be mapped to a kinetic aggregation rate constant in this limit, which helps connect our descriptions directly with physical quantities used in thixotropic constitutive modeling. The details are shown in Supplementary Information, \S~3.

% subsection break -----------------------------------------------------

\subsection{Step down in shear rate for three model materials\label{subsec:results-fits-disc-down-others}}

\begin{figure}[!ht]
	\begin{tabular}{ccc}
		\begin{minipage}[!ht]{0.32\textwidth}
			\begin{center}
			    (a) carbon black
				\includegraphics[scale=0.25,trim={0 0 0 0},clip]{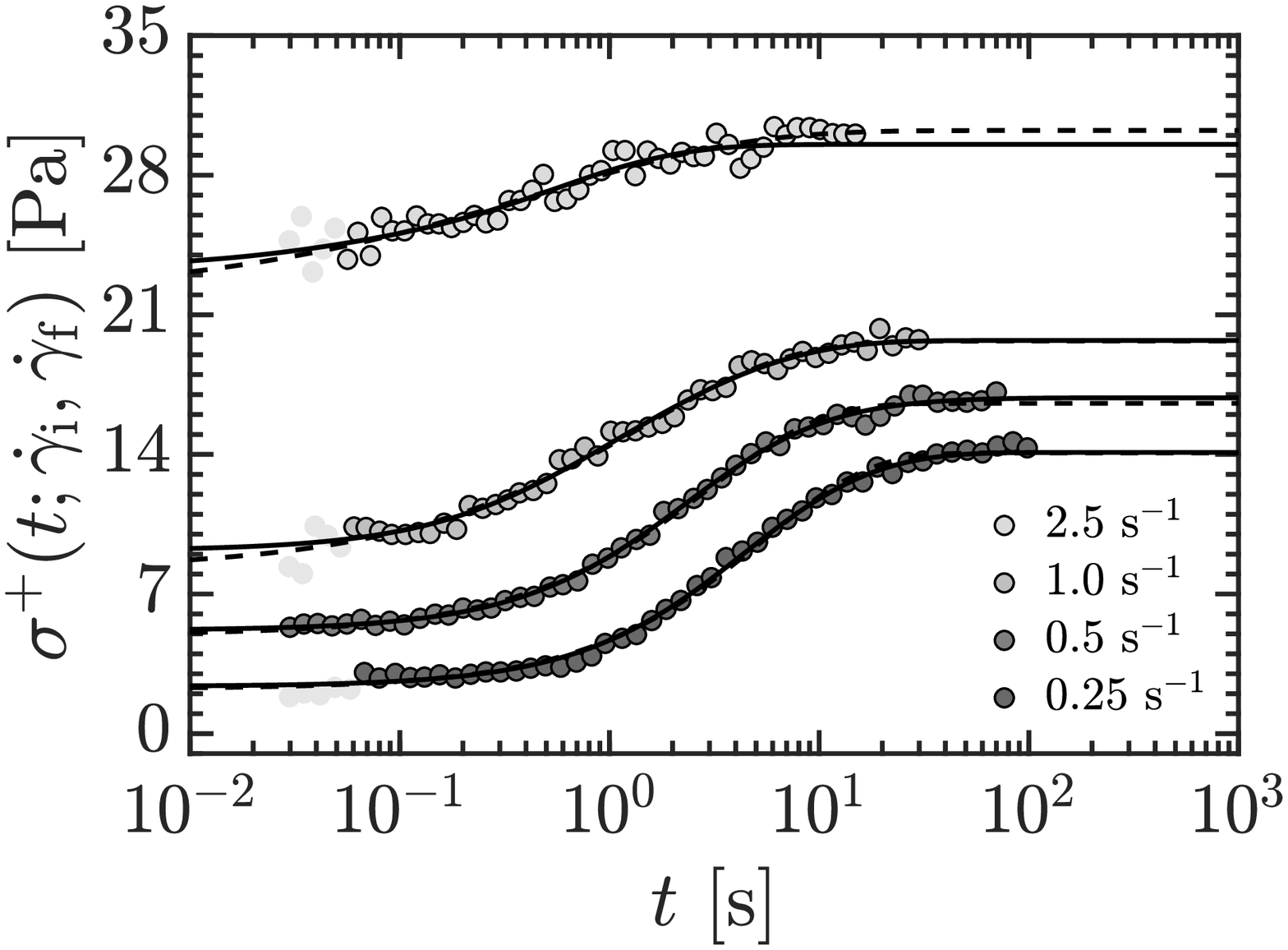}
			\end{center}
		\end{minipage}
		\begin{minipage}[!ht]{0.32\textwidth}
			\begin{center}
			    (b) fumed silica
				\includegraphics[scale=0.25,trim={0 0 0 0},clip]{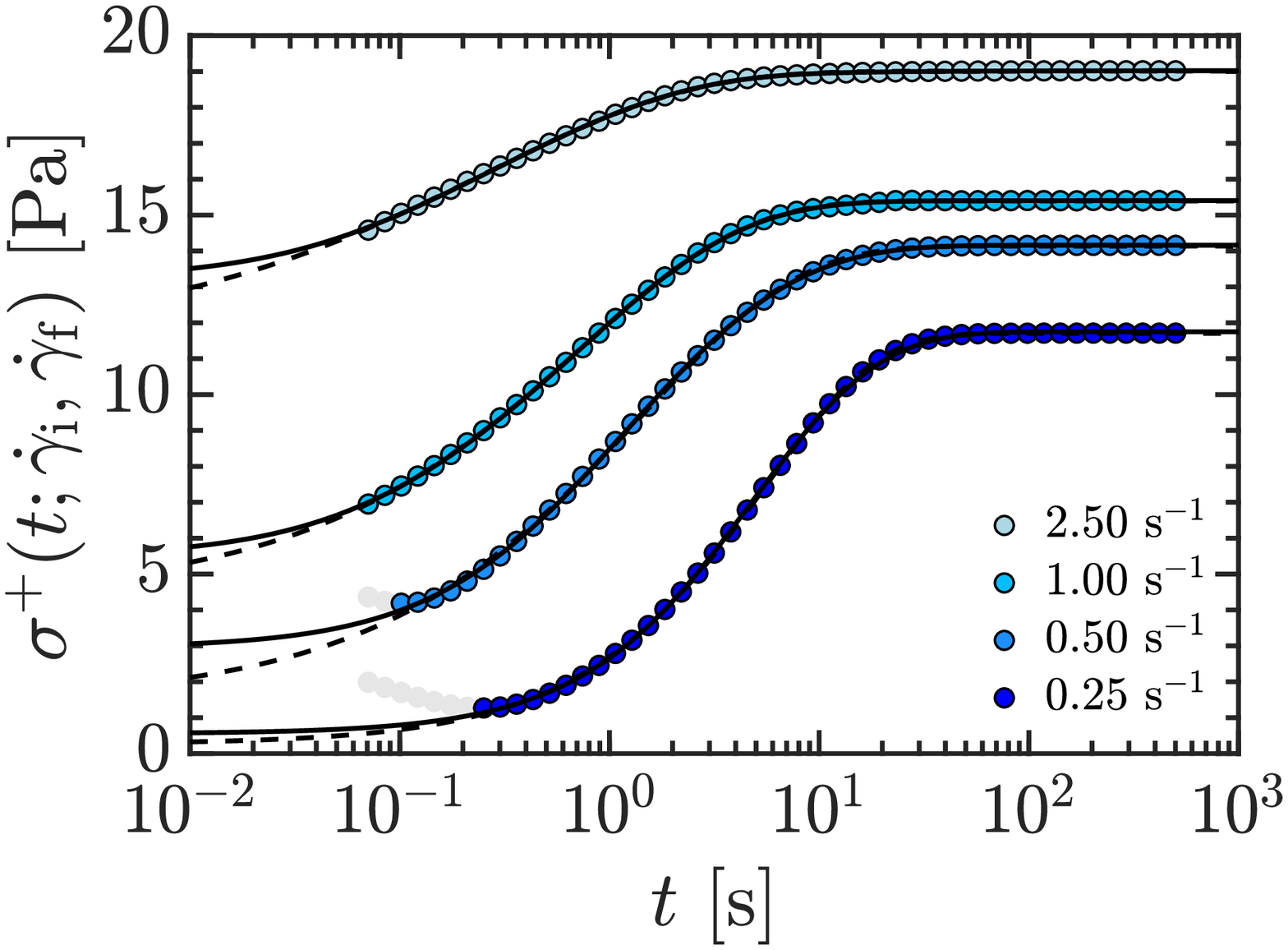}
			\end{center}
		\end{minipage}
		\begin{minipage}[!ht]{0.32\textwidth}
			\begin{center}
			    (c) Carbopol
				\includegraphics[scale=0.25,trim={0 0 0 0},clip]{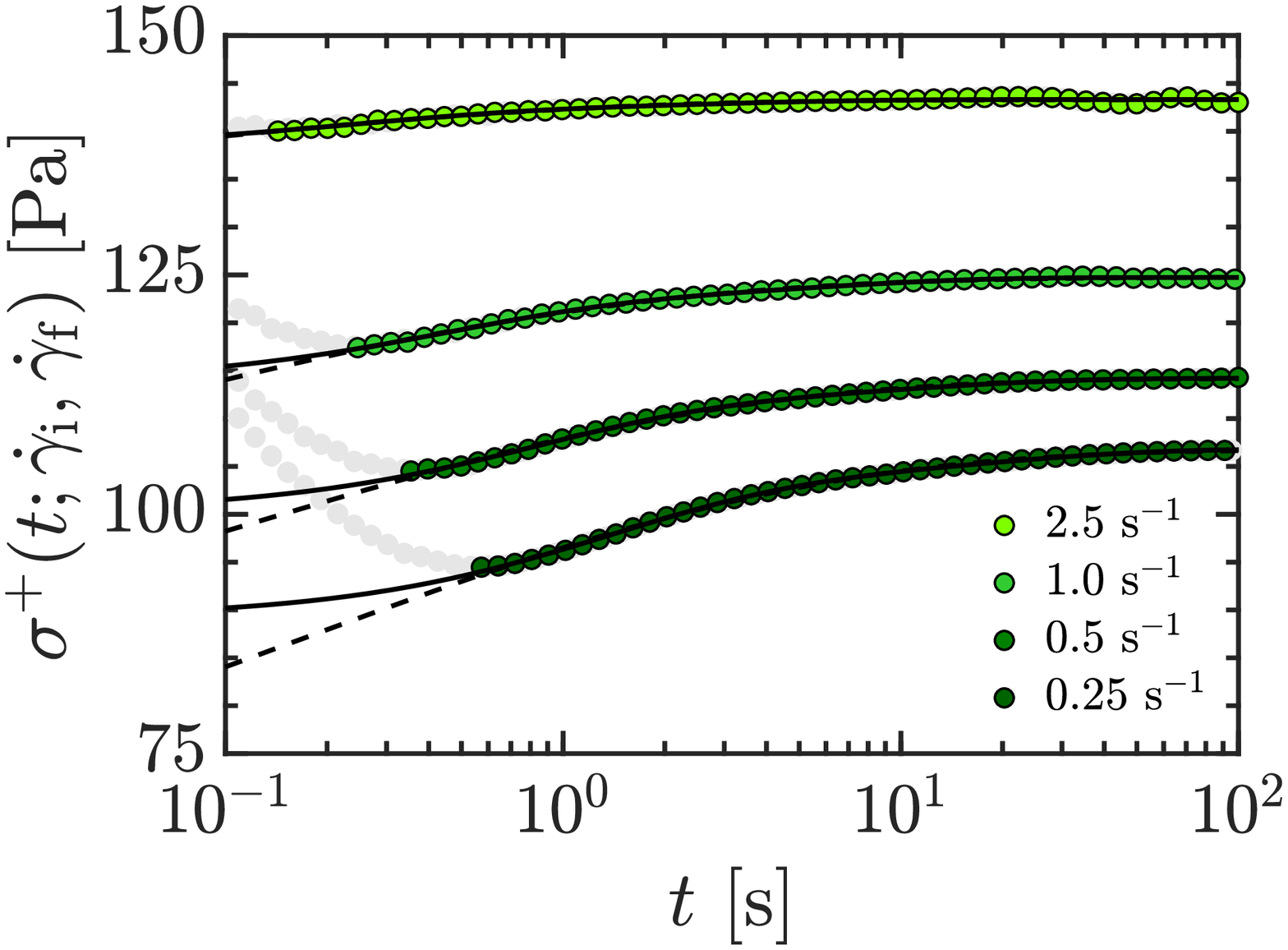}
			\end{center}
		\end{minipage}
	\end{tabular}
	\vspace{\floatsep}
	\begin{tabular}{ccc}
		\begin{minipage}[!ht]{0.32\textwidth}
			\begin{center}
				\includegraphics[scale=0.25,trim={0 0 0 0},clip]{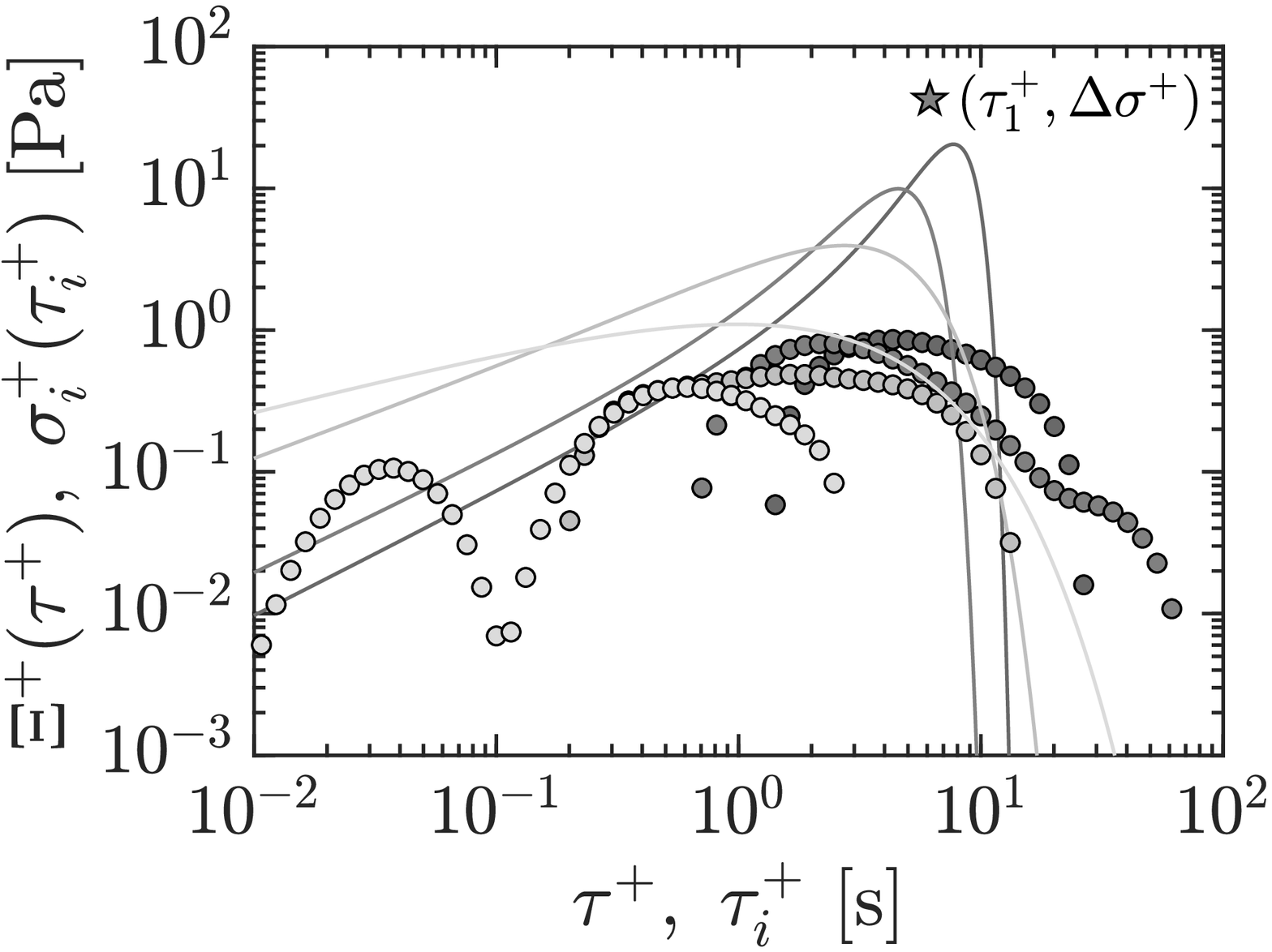}
			\end{center}
		\end{minipage}
		\begin{minipage}[!ht]{0.32\textwidth}
			\begin{center}
				\includegraphics[scale=0.25,trim={0 0 0 0},clip]{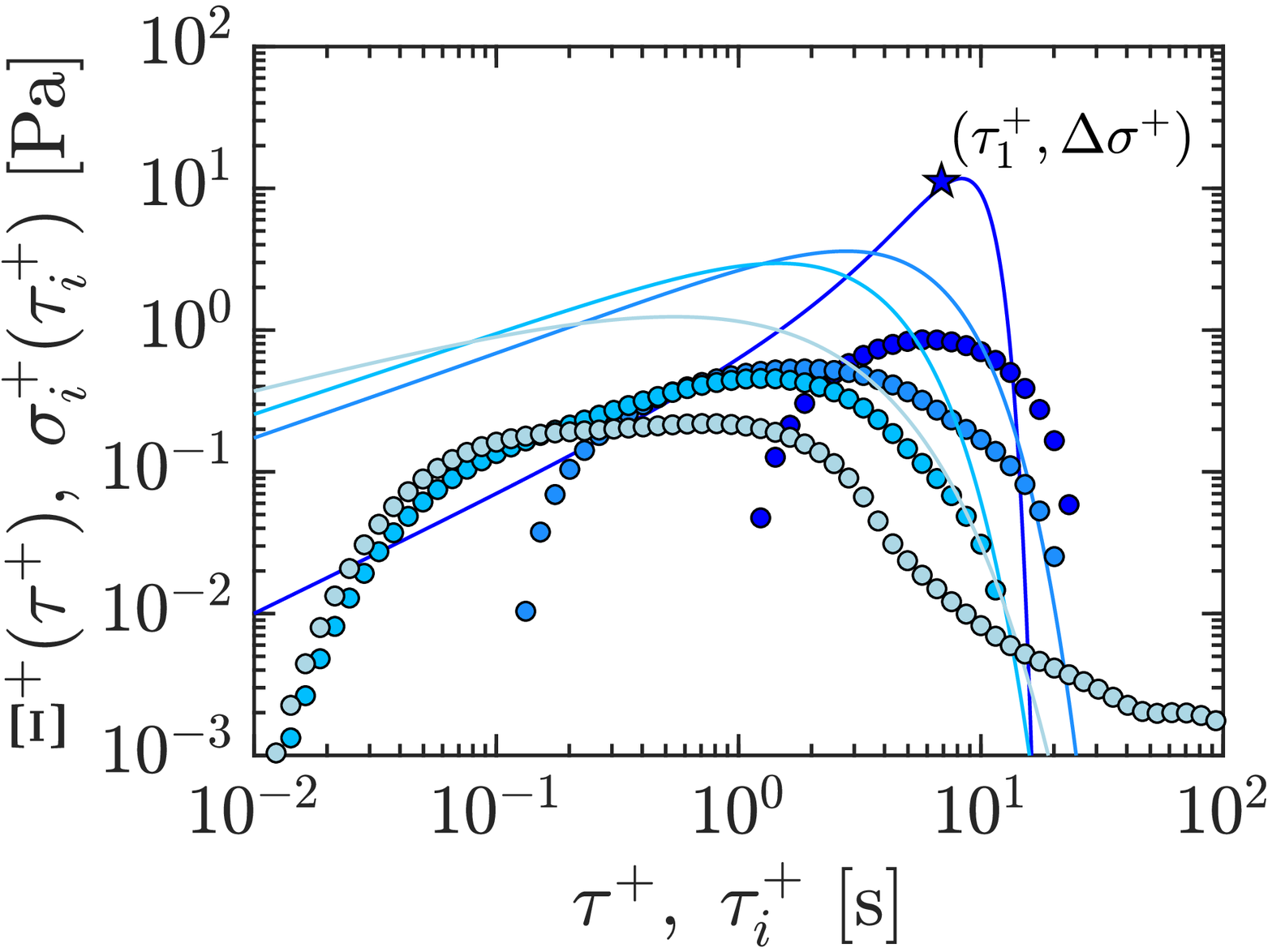}
			\end{center}
		\end{minipage}
		\begin{minipage}[!ht]{0.32\textwidth}
			\begin{center}
				\includegraphics[scale=0.25,trim={0 0 0 0},clip]{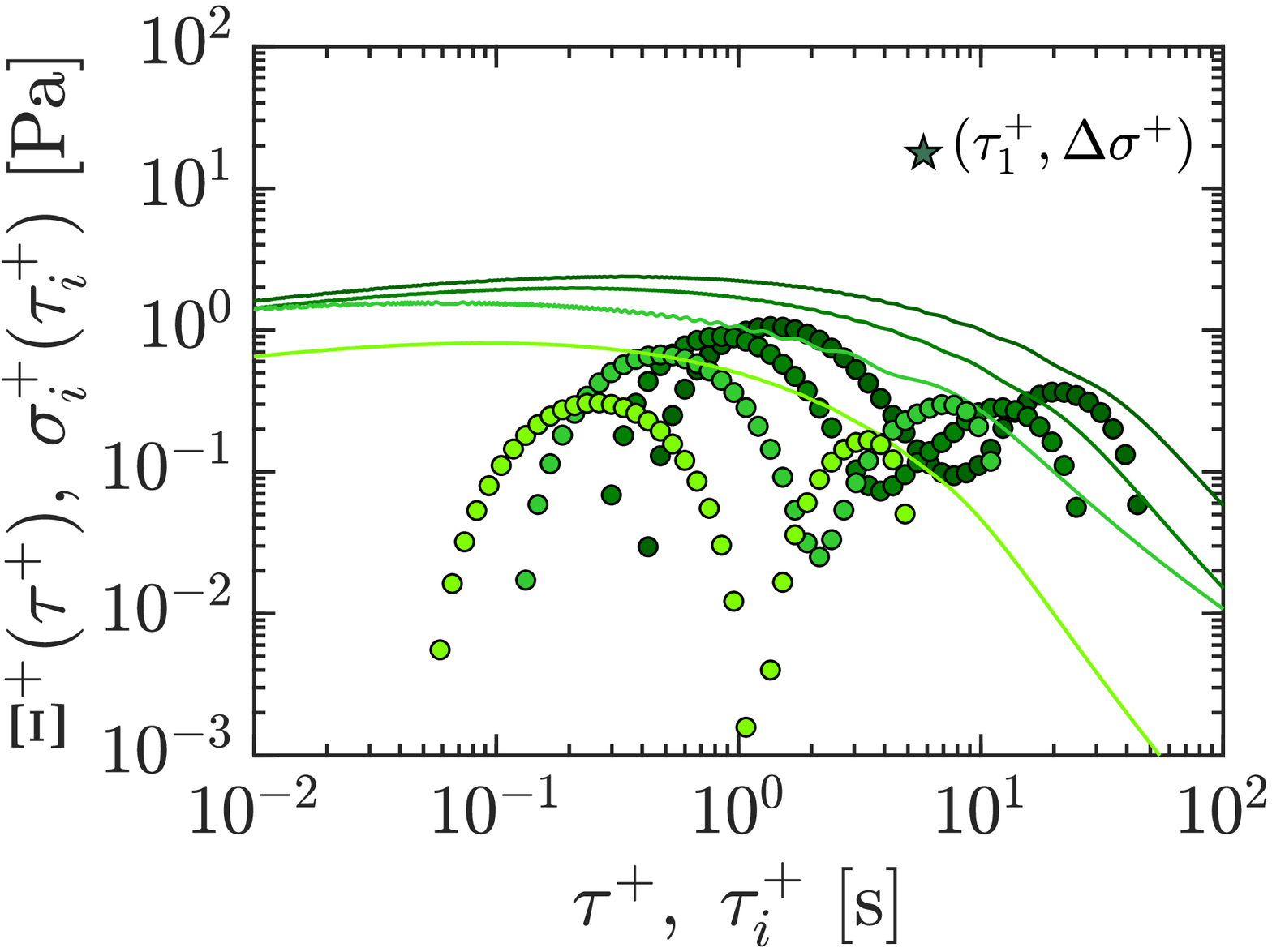}
			\end{center}
		\end{minipage}
	\end{tabular}
\caption{\label{fig:down-disc-others}Step down in shear data for (a) 3.23~vol\% carbon black (from Dullaert \emph{et al.} \cite{DullaertMewis_structkinetics2006}), (b) 2.9~vol\% fumed silica (from Armstrong \emph{et al.} \cite{ArmWag_JOR2016}), (c) 1~wt\% Carbopol. The top row shows the transient stress in recovery for each material, the fit lines correspond to the discrete (solid lines) and stretched exponential continuous (dashed lines) spectra. The bottom row shows the corresponding discrete (circles) and stretched exponential continuous (solid lines) spectra, similar to the data shown for Laponite in Fig.~\ref{fig:disc_down_stress}. The $\bigstar$ in each spectrum plot shows the summarizing metric for the most extreme case of thixotropy, corresponding to $\dot{\gamma}_{\rm f}=0.25~\rm{s}^{-1}$.}
\end{figure}

In addition to Laponite, we analyzed thixotropic recovery data for three other materials listed in \S~\ref{subsec:materials}. Fig.~\ref{fig:down-disc-others} shows the data for each material as separate columns, with the transient stress data and fit spectra for each, similar to Fig.~\ref{fig:disc_down_stress}. The features observed for both the stretched exponential $\Xi^+$ and discrete $\sigma^+_i$ in recovery for Laponite are similarly observed for these materials. The thixotropic timescales range from $<1~{\rm s}$ (e.g.\ carbon black at $\dot{\gamma}_{\rm f}=2.5~\rm{s}^{-1}$) to $>10~{\rm s}$ (e.g. fumed silica at $\dot{\gamma}_{\rm f}=0.25~\rm{s}^{-1}$). For carbon black and silica, the amount of stress change relative to the steady state value increases as $\dot{\gamma}_{\rm f}$ decreases, while this ratio is much smaller for Carbopol at all shear rates; this is quantified with $\Delta\sigma^+/\sigma_{\infty}$, shown in Fig.~\ref{fig:Ashby_all}. We also observe the existence of distinct peaks in the spectra, in addition to a spread in timescales, for Laponite, carbon black, and Carbopol, whereas we observe a broad spectrum and an absence of distinct, separate peaks in the fumed silica data. The polydispersity of thixotropic modes is captured in the spectral method developed here, including the ability to see distinct groups of modes via peaks in the distributions. The reduced summarizing metrics for Fig.~\ref{fig:disc_down_stress}, \ref{fig:down-disc-others} are shown in the following section.

%%%%%%%%%%%%%%%%%%%%%%%%%%%%%%%%%%%%%%%%%%%%%%%%%%%%%%%%%%%%%%%%%%%%%%%%%%%%%%%%
%===============================================================================
% section break
%===============================================================================
%%%%%%%%%%%%%%%%%%%%%%%%%%%%%%%%%%%%%%%%%%%%%%%%%%%%%%%%%%%%%%%%%%%%%%%%%%%%%%%%

\section{Low-dimensional descriptions from spectra\label{sec:lowdim}}
\begin{figure}[!ht]
	\begin{minipage}[!ht]{0.49\textwidth}
		(a)
		\centering
		\includegraphics[scale=0.39,trim={0 0 0 0},clip]{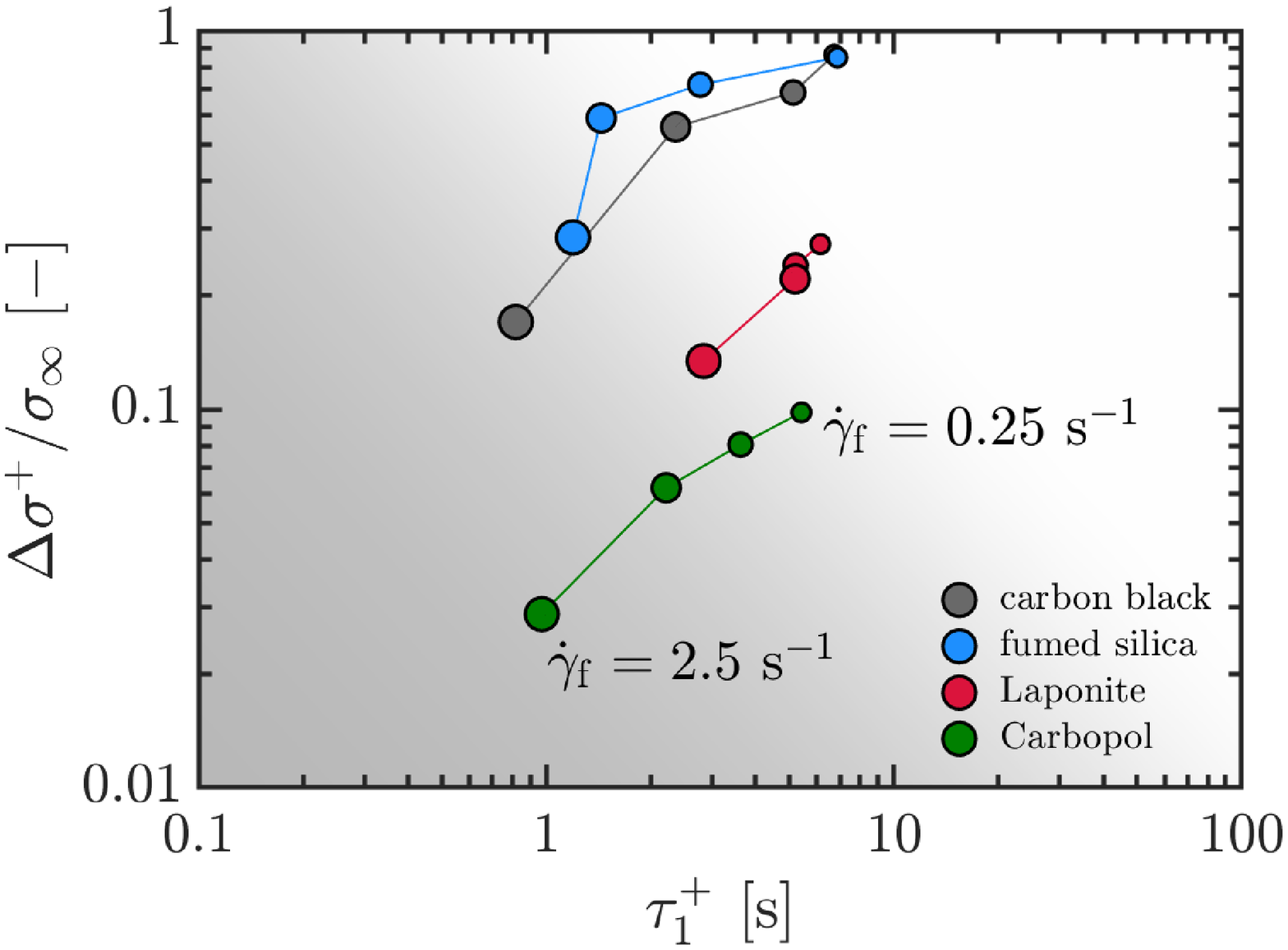}
	\end{minipage}
	\begin{minipage}[!ht]{0.49\textwidth}
		(b)
		\centering
		\includegraphics[scale=0.39,trim={0 0 0 0},clip]{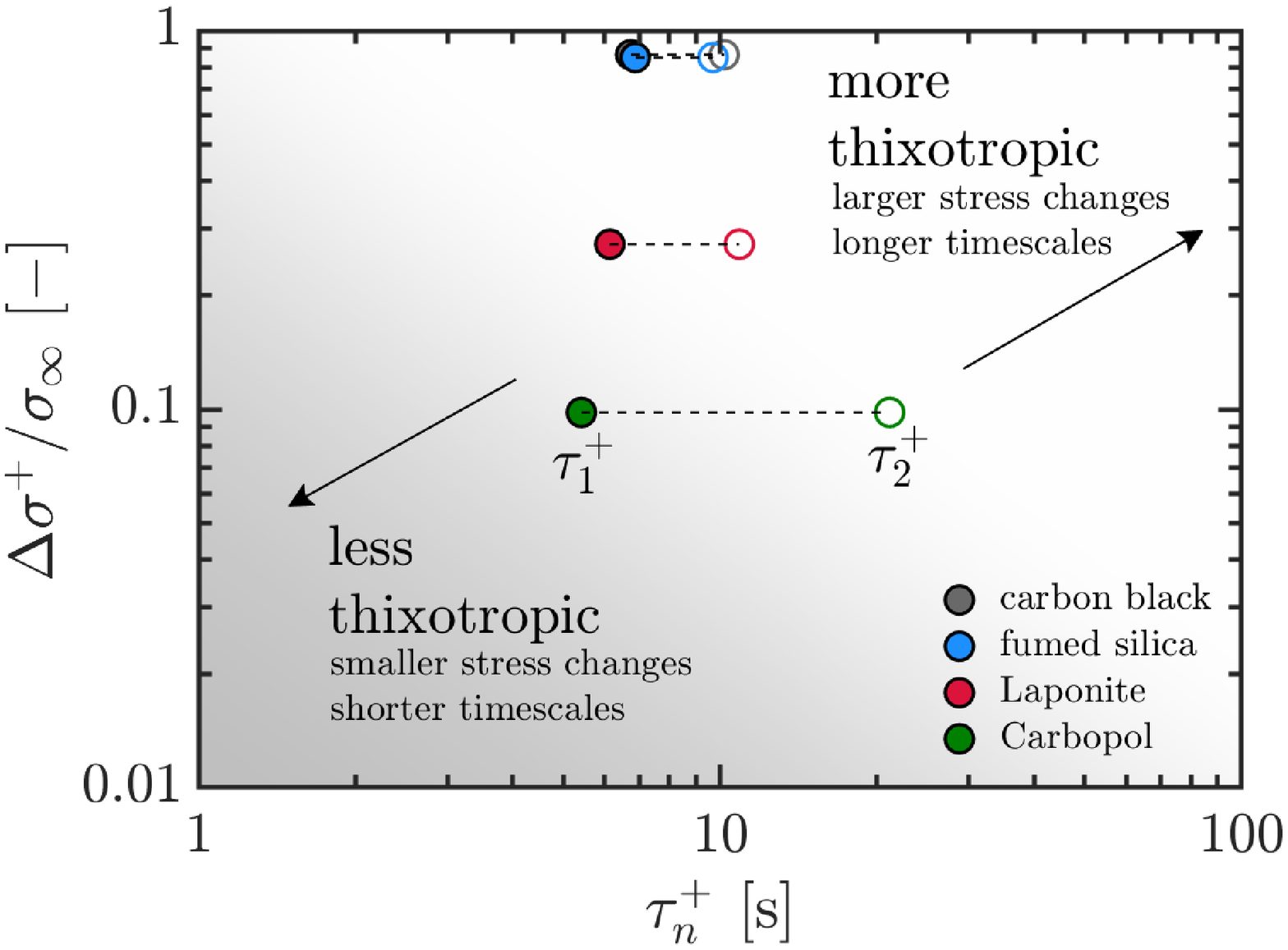}
	\end{minipage}
	\caption{\label{fig:Ashby_all}Ashby co-plots for thixotropic recovery, based on data in Figs.~\ref{fig:disc_down_stress}, \ref{fig:down-disc-others}. (a) Reduced metrics $(\tau^+_1,\Delta\sigma^+/\sigma_\infty)$ are obtained from discrete spectrum fits to step down in shear data for different fluids. The symbols get smaller as the final shear rate decreases ($\dot{\gamma}_{\rm f} = 2.50,~1.00,~0.50,~0.25~{\rm s}^{-1}$) for each material. (b) The mean timescales $\tau^+_1,\tau^+_2$ (visualizing ${\rm PDI}=\tau^+_2/\tau^+_1$ for the most thixotropic response for each material (observed at $\dot{\gamma}_{\rm f} = 0.25~{\rm s}^{-1}$) plotted against $\Delta\sigma^+/\sigma_\infty$.}
\end{figure}

Fig.~\ref{fig:Ashby_all} is an Ashby-style co-plot \cite{Ashby_book2011} of reduced, low-dimensional metrics of thixotropic recovery spectra from Figs.~\ref{fig:disc_down_stress}~and~\ref{fig:down-disc-others}. This allows direct comparison between the four different materials using $\Delta\sigma^+/\sigma_{\infty}$ \emph{versus} $\tau^+_1$, where $\Delta\sigma^+$ is the total stress recovered under $\dot{\gamma}_{\rm f}$, and is obtained from Eq.~\ref{eq:moments-M0-deltasigma}. The recovered stress is normalized by the steady state stress at long times, defined as $\sigma_{\infty} \equiv \lim_{t\rightarrow\infty}\sigma(t)$. Since $\sigma_{\infty}$ is used as the reference stress and the recovery is quantified as a percentage of this value, we get
\begin{align}
	0 \leq \frac{\Delta\sigma^+}{\sigma_{\infty}} \leq 1,
\end{align}
and this allows for comparison of thixotropic recovery for different materials. $\tau^+_1$ is the first mean timescale of recovery, obtained by using $n=1$ in Eq.~\ref{eq:taun}.

The closer the value of $\Delta\sigma^+/\sigma_{\infty}$ is to 1, the greater is the change in stress due to thixotropic recovery relative to the final state of the sample, and the greater the significance of thixotropic effects in the material for a given set of $\left( \dot{\gamma}_{\rm i},\dot{\gamma}_{\rm f} \right)$. Among these data, there is a single value of $\dot{\gamma}_{\rm i} = 5~{\rm s}^{-1}$, and only $\dot{\gamma}_{\rm f}$ is varied. Accordingly, we see the effect of changing the shear history in Fig.~\ref{fig:Ashby_all}(a), where smaller symbols correspond to a lower $\dot{\gamma}_{\rm f}$, approaching the Brownian-dominated recovery limit. As $\dot{\gamma}_{\rm f}$ is decreased, $\Delta\sigma^+/\sigma_{\infty}$ increases, as observed in Fig.~\ref{fig:Ashby_all}, and hence the degree of thixotropy also increases in terms of the stress recovered. But as mentioned in \S~\ref{sec:intro}, one must consider both stresses and timescales to judge the degree of thixotropy in a material.

The longer the timescale $\tau^+_1$, the longer the thixotropic transients last, and the greater the significance of thixotropic effects in the material, once again for a given set of $\left( \dot{\gamma}_{\rm i},\dot{\gamma}_{\rm f} \right)$. The same trend with changing $\dot{\gamma}_{\rm f}$ is observed in $\tau^+_1$ as with $\Delta\sigma^+/\sigma_{\infty}$: as $\dot{\gamma}_{\rm f}$ is decreased, $\tau^+_1$ increases, and thus the degree of thixotropy is most significant at the lowest $\dot{\gamma}_{\rm f}$ (Fig.~\ref{fig:disc_down_stress}, Fig.~\ref{fig:down-disc-others}). This is observed for all four materials shown in Fig.~\ref{fig:Ashby_all}(a). We see that the timescales are comparable across all materials ($\approx 1-10$~s), but the fractional stress recovery is very different. Considering the lowest $\dot{\gamma}_{\rm f}$ for each, Carbopol has $\Delta\sigma^+/\sigma_{\infty} = 0.09$, and is therefore the least thixotropic, even though its $\tau^+_1 = 5.41$~s is similar to that of the other materials. Laponite is more thixotropic than Carbopol, while carbon black and fumed silica are the most thixotropic.

The trends in Fig.~\ref{fig:Ashby_all}(a) form a compact summary and are consistent with basic thixotropic constitutive models. As mentioned in \S~\ref{sec:results-fits-disc}, the monotonic trend in $\tau^+_1$ with $\dot{\gamma}_{\rm f}$ is in accordance to classical structural kinetics-based thixotropic models in the literature \cite{MewisWagnerReview2009,LarsonWei-JoR2019} (see Supplementary Material, \S~3). Structure recovery through Brownian motion proceeds slowly, and at lower $\dot{\gamma}_{\rm f}$, the effect of Brownian dynamics is more prominent. The increase in $\Delta\sigma^+/\sigma_{\infty}$ with decreasing $\dot{\gamma}_{\rm f}$ also is physically meaningful; as $\dot{\gamma}_{\rm f}$ decreases, $\rm Pe$ decreases, and Brownian dynamics dominate over advective hydrodynamics, and more structure can build up (or equivalently, less structure is broken down due to shear-induced stresses). It is interesting that the two quantities used here qualitatively depend on $\dot{\gamma}_{\rm f}$ in the same way. The Ashby plots are simple to interpret and use: the further away a material lies from the origin, the more significant is the effect of thixotropy in the material (moving from the gray to the white region).

Fig.~\ref{fig:Ashby_all}(a) does not indicate the breadth of underlying thixotropic spectra, but showing both $\tau^+_1$ and $\tau^+_2$, as in Fig.~\ref{fig:Ashby_all}(b) for the minimum $\dot{\gamma}_{\rm f}$ data, reveals this. The more distinct $\tau^+_1$ and $\tau^+_2$ are, the more polydisperse the thixotropic recovery spectrum, ${\rm PDI} \equiv \tau^+_2/\tau^+_1$ from Eq.~\ref{eq:moments-tau-define-PDI} (${\rm PDI} = 3.31,~1.77,~1.49,~1.41$ for Carbopol, Laponite, carbon black, and fumed silica respectively). This representation in Fig.~\ref{fig:Ashby_all}(b) is general, a description applicable to any material, even those with different underlying constitutive behavior. It is therefore a powerful tool that is useful for comparison and selection across different microstructural classes of thixotropic materials. This technique has already been applied by the authors in studying drop impact dynamics of thixotropic yield-stress fluids \cite{SSRHE_JFM2020,SSRHE_PRF2021}, where reduced thixotropic metrics obtained from step shear data were used to select a material (Laponite) with a long recovery timescale and significant stress changes which has important consequences in their drop impact stick-splash outcomes \cite{SSRHE_PRF2021}.

%%%%%%%%%%%%%%%%%%%%%%%%%%%%%%%%%%%%%%%%%%%%%%%%%%%%%%%%%%%%%%%%%%%%%%%%%%%%%%%%
%===============================================================================
% section break
%===============================================================================
%%%%%%%%%%%%%%%%%%%%%%%%%%%%%%%%%%%%%%%%%%%%%%%%%%%%%%%%%%%%%%%%%%%%%%%%%%%%%%%%

\section{Conclusions\label{sec:conclusions}}

In this work, we have developed a general framework for thixotropic spectra to quantitatively describe thixotropy using a distribution of recovery/breakdown timescales in step tests. The moments and means of the distributions are used to generate low-dimensional metrics of quantifying the degree of thixotropy in the material, and these descriptions are both model-independent and material agnostic. Alternative reduced metrics for quantifying thixotropic timescales have been used by Divoux \emph{et al.} \cite{Divoux_PRL2013} and Jamali \emph{et al.} \cite{Jamali_PRL2019} using areas of thixotropic hysteresis loops. However, the methods developed in our work employ faster experimental protocols, largely isolate the effect of viscoelastic transients from thixotropic ones, can be applied to not only unidirectional but also oscillatory shear tests, and most importantly, separately describe recovery and breakdown and reveal spectral distributions. We mention the analogy of the methods developed here with viscoelastic relaxation spectra in Table~\ref{ch2:tab:thixo-VE-compare}. Similarly, an analogy may be drawn between stress controlled step tests where the strain rate response is expressed analogous to retardation spectra distributed over retardation timescales.

The methodology proposed in this work is not free from limitations, experimental or analytical. Rheometer motors that impose strains/strain rates on samples via angular displacements have a finite response time to ramp up or down $(t_{\rm mot})$, and step changes are not instantaneous. The shortest duration of change between $\dot{\gamma}_{\rm i}$ and $\dot{\gamma}_{\rm f}$ is limited by the inertia and response time of the rotor-geometry assembly. The quickest step changes are essentially ramps that at best occur over $t_{\rm mot} \sim \mathcal{O}(10)$~ms, and therefore transients in $\sigma(t;\dot{\gamma}_{\rm i},\dot{\gamma}_{\rm f})$ for $t \lesssim t_{\rm mot}$ cannot be observed. The credibility of short-time data is also limited by viscoelastic waves in the sample \cite{RHE_baddata}. Consequently, the range of $\tau$ and $\tau_i$ used for fitting spectra to stress data has to be truncated. The temporal location and value of $\sigma_0$ is thus also influenced by $t_{\rm mot}$. This in turn makes the value of $\Delta\sigma/\sigma_{\infty}$ and therefore the reduced descriptions used in the Ashby charts susceptible to instrument limitations used to collect the data.

Specific details of the shape of the distribution are lost upon reduction into its moments. It is possible to generate completely different distributions and stress responses from the same set of moments, one prominent example being Anscombe's quartet \cite{Anscombe}. This also typifies the ill-posed nature of the fitting problem. Multiple distributions can give the same stress response, and therefore the same means and moments. Inverting the summation in Eq.~\ref{eq:discspec-intro} to obtain $\sigma_i$ is an ill-posed problem, similar to those faced with viscoelastic spectra. The approach of using a specific analytically defined continuous spectrum to fit to stress data and obtain $\Xi(\tau)$ may be a workaround, especially when employed with Bayesian inference methods and credibility metrics, as has been applied to LVE \cite{FreundEwoldt}, MAOS \cite{LucaRHEMAOS_2018}, and thixotropic constitutive model selection \cite{ArmWag_JOR2016}.

Future work will benefit from this methodology of thixotropic spectra and Ashby diagrams. The high-density discrete spectra $\sigma_i$ can be employed to obtain dominant timescale modes, and a low-density superposition of continuous spectra centered around these timescales could be used to fit the data to very high accuracy. The method developed here is not predictive, but only descriptive. Yet, the value of descriptive methods cannot be overstated, e.g.\ material properties which are extremely useful in rheological analysis despite not being predictive. Our method also gives insight into thixotropic properties of fluids that might inform predictive models; for instance, revealing or suggesting shapes of underlying spectra for multi-mode constitutive equations, which can lead to better predictions. Shapes of the discrete spectra can also be used to inform the shape of parameterized continuous distributions of $\Xi(\tau)$. These make the fitting process less computationally intensive. The concept of thixotropic spectra also provides interpretation to commonly-used fit functions, such as the stretched exponential (Eq.~\ref{eq:strexp_stress},~\ref{eq:strexp_spectrum}). Lastly, the methodology developed here can be applied to any material system that evolves with time to obtain summarizing metrics that describe the data in a simple yet meaningful way. We now have a way to judge ``how thixotropic'' a material is in terms of both timescale and amount of change, which is important for dimensionless groups involving thixotropic and aging systems in general (e.g.\ the Mutation number \cite{Winter_1994,Winter_2018,GeriGHM_PRX2018}), and for rheological design requirements involving thixotropy \cite{ChaiRHE_ARFM} in applications such as 3D printing \cite{Poole_SM2019}, spray coating, fire suppression \cite{BCB_GFM2015,SSRHE_GFM2018}, flow batteries \cite{Helal_2014,Helal_2016,Narayanan}, food processing, and other uses not yet imagined.

%%%%%%%%%%%%%%%%%%%%%%%%%%%%%%%%%%%%%%%%%%%%%%%%%%%%%%%%%%%%%%%%%%%%%%%%%%%%%%%%
%===============================================================================
% section break
%===============================================================================
%%%%%%%%%%%%%%%%%%%%%%%%%%%%%%%%%%%%%%%%%%%%%%%%%%%%%%%%%%%%%%%%%%%%%%%%%%%%%%%%

\section{Supplementary Material\label{sec:SI}}
See Supplementary Information for more information about the continuous spectrum for the stretched exponential and the parameters for the step down data fits for all materials to the stretched exponential model.

%%%%%%%%%%%%%%%%%%%%%%%%%%%%%%%%%%%%%%%%%%%%%%%%%%%%%%%%%%%%%%%%%%%%%%%%%%%%%%%%
%===============================================================================
% section break
%===============================================================================
%%%%%%%%%%%%%%%%%%%%%%%%%%%%%%%%%%%%%%%%%%%%%%%%%%%%%%%%%%%%%%%%%%%%%%%%%%%%%%%%

\section{Acknowledgements\label{sec:acknowledgements}}
This project was funded by the National Science Foundation CAREER Award, CBET-1351342, and the Joint Center for Energy Storage Research (JCESR), U.S.\ Department of Energy. S.S.\ thanks Y.\ Wang, N.\ Ramlawi, and Dr.\ C.\ Saengow for helpful discussions.

%%%%%%%%%%%%%%%%%%%%%%%%%%%%%%%%%%%%%%%%%%%%%%%%%%%%%%%%%%%%%%%%%%%%%%%%%%%%%%%%
%===============================================================================
% section break
%===============================================================================
%%%%%%%%%%%%%%%%%%%%%%%%%%%%%%%%%%%%%%%%%%%%%%%%%%%%%%%%%%%%%%%%%%%%%%%%%%%%%%%%

\bibliography{Thixo-timescales}

%aipnum4-2.bst 2019-01-14 (MD) hand-edited version of apsrev4-1.bst
\providecommand{\noopsort}[1]{}\providecommand{\singleletter}[1]{#1}\newcommand{\noop}[1]{}
\begin{thebibliography}{56}%
\makeatletter
\providecommand \@ifxundefined [1]{%
 \@ifx{#1\undefined}
}%
\providecommand \@ifnum [1]{%
 \ifnum #1\expandafter \@firstoftwo
 \else \expandafter \@secondoftwo
 \fi
}%
\providecommand \@ifx [1]{%
 \ifx #1\expandafter \@firstoftwo
 \else \expandafter \@secondoftwo
 \fi
}%
\providecommand \natexlab [1]{#1}%
\providecommand \enquote  [1]{``#1''}%
\providecommand \bibnamefont  [1]{#1}%
\providecommand \bibfnamefont [1]{#1}%
\providecommand \citenamefont [1]{#1}%
\providecommand \href@noop [0]{\@secondoftwo}%
\providecommand \href [0]{\begingroup \@sanitize@url \@href}%
\providecommand \@href[1]{\@@startlink{#1}\@@href}%
\providecommand \@@href[1]{\endgroup#1\@@endlink}%
\providecommand \@sanitize@url [0]{\catcode `\\12\catcode `\$12\catcode
  `\&12\catcode `\#12\catcode `\^12\catcode `\_12\catcode `\%12\relax}%
\providecommand \@@startlink[1]{}%
\providecommand \@@endlink[0]{}%
\providecommand \url  [0]{\begingroup\@sanitize@url \@url }%
\providecommand \@url [1]{\endgroup\@href {#1}{\urlprefix }}%
\providecommand \urlprefix  [0]{URL }%
\providecommand \Eprint [0]{\href }%
\providecommand \doibase [0]{https://doi.org/}%
\providecommand \selectlanguage [0]{\@gobble}%
\providecommand \bibinfo  [0]{\@secondoftwo}%
\providecommand \bibfield  [0]{\@secondoftwo}%
\providecommand \translation [1]{[#1]}%
\providecommand \BibitemOpen [0]{}%
\providecommand \bibitemStop [0]{}%
\providecommand \bibitemNoStop [0]{.\EOS\space}%
\providecommand \EOS [0]{\spacefactor3000\relax}%
\providecommand \BibitemShut  [1]{\csname bibitem#1\endcsname}%
\let\auto@bib@innerbib\@empty
%</preamble>
\bibitem [{\citenamefont {Fan}\ \emph {et~al.}(2014)\citenamefont {Fan},
  \citenamefont {Woodford}, \citenamefont {Li}, \citenamefont {Baram},
  \citenamefont {Smith}, \citenamefont {Helal}, \citenamefont {McKinley},
  \citenamefont {Carter},\ and\ \citenamefont {Chiang}}]{Helal_2014}%
  \BibitemOpen
  \bibfield  {author} {\bibinfo {author} {\bibfnamefont {F.~Y.}\ \bibnamefont
  {Fan}}, \bibinfo {author} {\bibfnamefont {W.~H.}\ \bibnamefont {Woodford}},
  \bibinfo {author} {\bibfnamefont {Z.}~\bibnamefont {Li}}, \bibinfo {author}
  {\bibfnamefont {N.}~\bibnamefont {Baram}}, \bibinfo {author} {\bibfnamefont
  {K.~C.}\ \bibnamefont {Smith}}, \bibinfo {author} {\bibfnamefont
  {A.}~\bibnamefont {Helal}}, \bibinfo {author} {\bibfnamefont {G.~H.}\
  \bibnamefont {McKinley}}, \bibinfo {author} {\bibfnamefont {W.~C.}\
  \bibnamefont {Carter}},\ and\ \bibinfo {author} {\bibfnamefont {Y.~M.}\
  \bibnamefont {Chiang}},\ }\bibfield  {title} {\enquote {\bibinfo {title}
  {Polysulfide flow batteries enabled by percolating nanoscale conductor
  networks},}\ }\href@noop {} {\bibfield  {journal} {\bibinfo  {journal} {Nano
  Lett.}\ }\textbf {\bibinfo {volume} {14}},\ \bibinfo {pages} {2210--2218}
  (\bibinfo {year} {2014})}\BibitemShut {NoStop}%
\bibitem [{\citenamefont {Helal}, \citenamefont {Divoux},\ and\ \citenamefont
  {McKinley}(2016)}]{Helal_2016}%
  \BibitemOpen
  \bibfield  {author} {\bibinfo {author} {\bibfnamefont {A.}~\bibnamefont
  {Helal}}, \bibinfo {author} {\bibfnamefont {T.}~\bibnamefont {Divoux}},\ and\
  \bibinfo {author} {\bibfnamefont {G.~H.}\ \bibnamefont {McKinley}},\
  }\bibfield  {title} {\enquote {\bibinfo {title} {Simultaneous rheoelectric
  measurements of strongly conductive complex fluids},}\ }\href@noop {}
  {\bibfield  {journal} {\bibinfo  {journal} {Phys. Rev. Appl.}\ }\textbf
  {\bibinfo {volume} {6}},\ \bibinfo {pages} {1--19} (\bibinfo {year}
  {2016})}\BibitemShut {NoStop}%
\bibitem [{\citenamefont {Narayanan}, \citenamefont {Mugele},\ and\
  \citenamefont {Duits}(2017)}]{Narayanan}%
  \BibitemOpen
  \bibfield  {author} {\bibinfo {author} {\bibfnamefont {A.}~\bibnamefont
  {Narayanan}}, \bibinfo {author} {\bibfnamefont {F.}~\bibnamefont {Mugele}},\
  and\ \bibinfo {author} {\bibfnamefont {M.~H.~G.}\ \bibnamefont {Duits}},\
  }\bibfield  {title} {\enquote {\bibinfo {title} {Mechanical history
  dependence in carbon black suspensions for flow batteries: A rheo-impedance
  study},}\ }\href@noop {} {\bibfield  {journal} {\bibinfo  {journal}
  {Langmuir}\ }\textbf {\bibinfo {volume} {33}},\ \bibinfo {pages} {1629--1638}
  (\bibinfo {year} {2017})}\BibitemShut {NoStop}%
\bibitem [{\citenamefont {Wang}\ and\ \citenamefont {Ewoldt}()}]{Wang_JoR}%
  \BibitemOpen
  \bibfield  {author} {\bibinfo {author} {\bibfnamefont {Y.}~\bibnamefont
  {Wang}}\ and\ \bibinfo {author} {\bibfnamefont {R.~H.}\ \bibnamefont
  {Ewoldt}},\ }\bibfield  {title} {\enquote {\bibinfo {title} {New insights on
  carbon black suspension rheology - anisotropic thixotropy and
  anti-thixotropy},}\ }\href@noop {} {\bibinfo  {journal} {Journal of Rheology,
  submitted}\ }\BibitemShut {NoStop}%
\bibitem [{\citenamefont {Dimitriou}\ and\ \citenamefont
  {Mckinley}(2014)}]{DimitriouMcKinley_SM2014}%
  \BibitemOpen
\bibfield  {journal} {  }\bibfield  {author} {\bibinfo {author} {\bibfnamefont
  {C.~J.}\ \bibnamefont {Dimitriou}}\ and\ \bibinfo {author} {\bibfnamefont
  {G.~H.}\ \bibnamefont {Mckinley}},\ }\bibfield  {title} {\enquote {\bibinfo
  {title} {A comprehensive constitutive law for waxy crude oil: A thixotropic
  yield stress fluid},}\ }\href@noop {} {\bibfield  {journal} {\bibinfo
  {journal} {Soft Matter}\ }\textbf {\bibinfo {volume} {10}},\ \bibinfo {pages}
  {6619} (\bibinfo {year} {2014})}\BibitemShut {NoStop}%
\bibitem [{\citenamefont {Glicerina}\ \emph {et~al.}(2015)\citenamefont
  {Glicerina}, \citenamefont {Balestra}, \citenamefont {Rosa},\ and\
  \citenamefont {Romani}}]{Glicerina2015}%
  \BibitemOpen
  \bibfield  {author} {\bibinfo {author} {\bibfnamefont {V.}~\bibnamefont
  {Glicerina}}, \bibinfo {author} {\bibfnamefont {F.}~\bibnamefont {Balestra}},
  \bibinfo {author} {\bibfnamefont {M.~D.}\ \bibnamefont {Rosa}},\ and\
  \bibinfo {author} {\bibfnamefont {S.}~\bibnamefont {Romani}},\ }\bibfield
  {title} {\enquote {\bibinfo {title} {Microstructural and rheological
  properties of white chocolate during processing},}\ }\href@noop {} {\bibfield
   {journal} {\bibinfo  {journal} {Food Bioprocess Technol.}\ }\textbf
  {\bibinfo {volume} {8}},\ \bibinfo {pages} {770--776} (\bibinfo {year}
  {2015})}\BibitemShut {NoStop}%
\bibitem [{\citenamefont {Jin}\ \emph {et~al.}(2011)\citenamefont {Jin},
  \citenamefont {Wu}, \citenamefont {Wei}, \citenamefont {Zhai},\ and\
  \citenamefont {Zhang}}]{Jin_Blood2011}%
  \BibitemOpen
  \bibfield  {author} {\bibinfo {author} {\bibfnamefont {L.}~\bibnamefont
  {Jin}}, \bibinfo {author} {\bibfnamefont {Z.}~\bibnamefont {Wu}}, \bibinfo
  {author} {\bibfnamefont {T.}~\bibnamefont {Wei}}, \bibinfo {author}
  {\bibfnamefont {J.}~\bibnamefont {Zhai}},\ and\ \bibinfo {author}
  {\bibfnamefont {X.}~\bibnamefont {Zhang}},\ }\bibfield  {title} {\enquote
  {\bibinfo {title} {Dye-sensitized solar cell based on blood mimetic
  thixotropy sol–gel electrolyte},}\ }\href@noop {} {\bibfield  {journal}
  {\bibinfo  {journal} {Chem. Commun.}\ }\textbf {\bibinfo {volume} {47}},\
  \bibinfo {pages} {997--999} (\bibinfo {year} {2011})}\BibitemShut {NoStop}%
\bibitem [{\citenamefont {Armstrong}\ and\ \citenamefont
  {Tussing}(2020)}]{Armstrong2020}%
  \BibitemOpen
  \bibfield  {author} {\bibinfo {author} {\bibfnamefont {M.~J.}\ \bibnamefont
  {Armstrong}}\ and\ \bibinfo {author} {\bibfnamefont {J.}~\bibnamefont
  {Tussing}},\ }\bibfield  {title} {\enquote {\bibinfo {title} {{A methodology
  for adding thixotropy to Oldroyd-8 family of viscoelastic models for
  characterization of human blood}},}\ }\href@noop {} {\bibfield  {journal}
  {\bibinfo  {journal} {Phys. Fluids}\ }\textbf {\bibinfo {volume} {32}},\
  \bibinfo {pages} {094111} (\bibinfo {year} {2020})}\BibitemShut {NoStop}%
\bibitem [{\citenamefont {Dullaert}\ and\ \citenamefont
  {Mewis}(2005)}]{DullaertMewis_ModelThixo2005}%
  \BibitemOpen
  \bibfield  {author} {\bibinfo {author} {\bibfnamefont {K.}~\bibnamefont
  {Dullaert}}\ and\ \bibinfo {author} {\bibfnamefont {J.}~\bibnamefont
  {Mewis}},\ }\bibfield  {title} {\enquote {\bibinfo {title} {A model system
  for thixotropy studies},}\ }\href@noop {} {\bibfield  {journal} {\bibinfo
  {journal} {Rheol. Acta}\ }\textbf {\bibinfo {volume} {45}},\ \bibinfo {pages}
  {23--32} (\bibinfo {year} {2005})}\BibitemShut {NoStop}%
\bibitem [{\citenamefont {Alessandrini}, \citenamefont {Lapasin},\ and\
  \citenamefont {Sturzi}(1982)}]{Alessandrini1982}%
  \BibitemOpen
  \bibfield  {author} {\bibinfo {author} {\bibfnamefont {A.}~\bibnamefont
  {Alessandrini}}, \bibinfo {author} {\bibfnamefont {R.}~\bibnamefont
  {Lapasin}},\ and\ \bibinfo {author} {\bibfnamefont {F.}~\bibnamefont
  {Sturzi}},\ }\bibfield  {title} {\enquote {\bibinfo {title} {The kinetics of
  thixotropic behaviour in clay/kaolin aqueous suspensions},}\ }\href@noop {}
  {\bibfield  {journal} {\bibinfo  {journal} {Chem. Eng. Commun.}\ }\textbf
  {\bibinfo {volume} {17}},\ \bibinfo {pages} {13--22} (\bibinfo {year}
  {1982})}\BibitemShut {NoStop}%
\bibitem [{\citenamefont {Beris}, \citenamefont {Stiakakis},\ and\
  \citenamefont {Vlassopoulos}(2008)}]{Beris_starJNNFM2008}%
  \BibitemOpen
  \bibfield  {author} {\bibinfo {author} {\bibfnamefont {A.~N.}\ \bibnamefont
  {Beris}}, \bibinfo {author} {\bibfnamefont {E.}~\bibnamefont {Stiakakis}},\
  and\ \bibinfo {author} {\bibfnamefont {D.}~\bibnamefont {Vlassopoulos}},\
  }\bibfield  {title} {\enquote {\bibinfo {title} {A thermodynamically
  consistent model for the thixotropic behavior of concentrated star polymer
  suspensions},}\ }\href@noop {} {\bibfield  {journal} {\bibinfo  {journal} {J.
  Non-Newtonian Fluid Mech.}\ }\textbf {\bibinfo {volume} {152}},\ \bibinfo
  {pages} {76--85} (\bibinfo {year} {2008})}\BibitemShut {NoStop}%
\bibitem [{\citenamefont {Burgos}, \citenamefont {Alexandrou},\ and\
  \citenamefont {Entov}(2001)}]{Burgos2001}%
  \BibitemOpen
  \bibfield  {author} {\bibinfo {author} {\bibfnamefont {G.~R.}\ \bibnamefont
  {Burgos}}, \bibinfo {author} {\bibfnamefont {A.~N.}\ \bibnamefont
  {Alexandrou}},\ and\ \bibinfo {author} {\bibfnamefont {V.}~\bibnamefont
  {Entov}},\ }\bibfield  {title} {\enquote {\bibinfo {title} {Thixotropic
  rheology of semisolid metal suspensions},}\ }\href@noop {} {\bibfield
  {journal} {\bibinfo  {journal} {J. Mater. Process. Technol.}\ }\textbf
  {\bibinfo {volume} {110}},\ \bibinfo {pages} {164--176} (\bibinfo {year}
  {2001})}\BibitemShut {NoStop}%
\bibitem [{\citenamefont {Kelessidis}(2008)}]{Kelessidis2008}%
  \BibitemOpen
  \bibfield  {author} {\bibinfo {author} {\bibfnamefont {V.}~\bibnamefont
  {Kelessidis}},\ }\bibfield  {title} {\enquote {\bibinfo {title}
  {Investigations on the thixotropy of bentonite suspensions},}\ }\href@noop {}
  {\bibfield  {journal} {\bibinfo  {journal} {Energy Sources A}\ }\textbf
  {\bibinfo {volume} {30}},\ \bibinfo {pages} {1729--1746} (\bibinfo {year}
  {2008})}\BibitemShut {NoStop}%
\bibitem [{\citenamefont {Mewis}(1979)}]{MewisReview1979}%
  \BibitemOpen
  \bibfield  {author} {\bibinfo {author} {\bibfnamefont {J.}~\bibnamefont
  {Mewis}},\ }\bibfield  {title} {\enquote {\bibinfo {title} {Thixotropy -- a
  general review},}\ }\href@noop {} {\bibfield  {journal} {\bibinfo  {journal}
  {J. Non-Newtonian Fluid Mech.}\ }\textbf {\bibinfo {volume} {6}},\ \bibinfo
  {pages} {1--20} (\bibinfo {year} {1979})}\BibitemShut {NoStop}%
\bibitem [{\citenamefont {Barnes}(1997)}]{Barnes_ThixoReview1997}%
  \BibitemOpen
  \bibfield  {author} {\bibinfo {author} {\bibfnamefont {H.~A.}\ \bibnamefont
  {Barnes}},\ }\bibfield  {title} {\enquote {\bibinfo {title} {Thixotropy - a
  review},}\ }\href@noop {} {\bibfield  {journal} {\bibinfo  {journal} {J.
  Non-Newtonian Fluid Mech.}\ }\textbf {\bibinfo {volume} {70}},\ \bibinfo
  {pages} {1--33} (\bibinfo {year} {1997})}\BibitemShut {NoStop}%
\bibitem [{\citenamefont {Mewis}\ and\ \citenamefont
  {Wagner}(2009)}]{MewisWagnerReview2009}%
  \BibitemOpen
  \bibfield  {author} {\bibinfo {author} {\bibfnamefont {J.}~\bibnamefont
  {Mewis}}\ and\ \bibinfo {author} {\bibfnamefont {N.~J.}\ \bibnamefont
  {Wagner}},\ }\bibfield  {title} {\enquote {\bibinfo {title} {Thixotropy},}\
  }\href@noop {} {\bibfield  {journal} {\bibinfo  {journal} {Adv. Colloid
  Interface Sci.}\ }\textbf {\bibinfo {volume} {147--148}},\ \bibinfo {pages}
  {214--227} (\bibinfo {year} {2009})}\BibitemShut {NoStop}%
\bibitem [{\citenamefont {Larson}\ and\ \citenamefont
  {Wei}(2019)}]{LarsonWei-JoR2019}%
  \BibitemOpen
  \bibfield  {author} {\bibinfo {author} {\bibfnamefont {R.~G.}\ \bibnamefont
  {Larson}}\ and\ \bibinfo {author} {\bibfnamefont {Y.}~\bibnamefont {Wei}},\
  }\bibfield  {title} {\enquote {\bibinfo {title} {A review of thixotropy and
  its rheological modeling},}\ }\href@noop {} {\bibfield  {journal} {\bibinfo
  {journal} {J. Rheol.}\ }\textbf {\bibinfo {volume} {63}},\ \bibinfo {pages}
  {477--501} (\bibinfo {year} {2019})}\BibitemShut {NoStop}%
\bibitem [{\citenamefont {Mewis}\ and\ \citenamefont
  {Wagner}(2012)}]{MewisWagner_book2012}%
  \BibitemOpen
  \bibfield  {author} {\bibinfo {author} {\bibfnamefont {J.}~\bibnamefont
  {Mewis}}\ and\ \bibinfo {author} {\bibfnamefont {N.~J.}\ \bibnamefont
  {Wagner}},\ }\href@noop {} {\emph {\bibinfo {title} {{Colloidal Suspension
  Rheology}}}}\ (\bibinfo  {publisher} {Cambridge University Press},\ \bibinfo
  {year} {2012})\BibitemShut {NoStop}%
\bibitem [{\citenamefont {Joshi}(2021)}]{Joshi_JOR2021}%
  \BibitemOpen
  \bibfield  {author} {\bibinfo {author} {\bibfnamefont {Y.~M.}\ \bibnamefont
  {Joshi}},\ }\bibfield  {title} {\enquote {\bibinfo {title} {{Thixotropy,
  nonmonotonic stress relaxation, and the second law of thermodynamics}},}\
  }\href@noop {} {\bibfield  {journal} {\bibinfo  {journal} {J. Rheol.}\
  }\textbf {\bibinfo {volume} {66}},\ \bibinfo {pages} {111--123} (\bibinfo
  {year} {2021})}\BibitemShut {NoStop}%
\bibitem [{\citenamefont {Wei}, \citenamefont {Solomon},\ and\ \citenamefont
  {Larson}(2016)}]{WeiSolomonLarson_SE-JOR2016}%
  \BibitemOpen
  \bibfield  {author} {\bibinfo {author} {\bibfnamefont {Y.}~\bibnamefont
  {Wei}}, \bibinfo {author} {\bibfnamefont {M.~J.}\ \bibnamefont {Solomon}},\
  and\ \bibinfo {author} {\bibfnamefont {R.~G.}\ \bibnamefont {Larson}},\
  }\bibfield  {title} {\enquote {\bibinfo {title} {Quantitative nonlinear
  thixotropic model with stretched exponential response in transient shear
  flows},}\ }\href@noop {} {\bibfield  {journal} {\bibinfo  {journal} {J.
  Rheol.}\ }\textbf {\bibinfo {volume} {60}},\ \bibinfo {pages} {1301}
  (\bibinfo {year} {2016})}\BibitemShut {NoStop}%
\bibitem [{\citenamefont {Larson}(2015)}]{Larson_ConstEq-JOR2015}%
  \BibitemOpen
  \bibfield  {author} {\bibinfo {author} {\bibfnamefont {R.~G.}\ \bibnamefont
  {Larson}},\ }\bibfield  {title} {\enquote {\bibinfo {title} {Constitutive
  equations for thixotropic fluids},}\ }\href@noop {} {\bibfield  {journal}
  {\bibinfo  {journal} {J. Rheol.}\ }\textbf {\bibinfo {volume} {59}},\
  \bibinfo {pages} {595} (\bibinfo {year} {2015})}\BibitemShut {NoStop}%
\bibitem [{\citenamefont {Tschoegl}(1989)}]{Tschoegl_book1989}%
  \BibitemOpen
  \bibfield  {author} {\bibinfo {author} {\bibfnamefont {N.~W.}\ \bibnamefont
  {Tschoegl}},\ }\href@noop {} {\emph {\bibinfo {title} {The Phenomenological
  Theory of Linear Viscoelastic Behavior}}}\ (\bibinfo  {publisher}
  {Springer-Verlag},\ \bibinfo {address} {Berlin},\ \bibinfo {year}
  {1989})\BibitemShut {NoStop}%
\bibitem [{\citenamefont {Vlad}\ \emph {et~al.}(1996)\citenamefont {Vlad},
  \citenamefont {Metzler}, \citenamefont {Nonnenmacher},\ and\ \citenamefont
  {Mackey}}]{Metzler1996}%
  \BibitemOpen
  \bibfield  {author} {\bibinfo {author} {\bibfnamefont {M.~O.}\ \bibnamefont
  {Vlad}}, \bibinfo {author} {\bibfnamefont {R.}~\bibnamefont {Metzler}},
  \bibinfo {author} {\bibfnamefont {T.~F.}\ \bibnamefont {Nonnenmacher}},\ and\
  \bibinfo {author} {\bibfnamefont {M.~C.}\ \bibnamefont {Mackey}},\ }\bibfield
   {title} {\enquote {\bibinfo {title} {Universality classes for asymptotic
  behavior of relaxation processes in systems with dynamical disorder:
  Dynamical generalizations of stretched exponential},}\ }\href@noop {}
  {\bibfield  {journal} {\bibinfo  {journal} {Energy Sources A}\ }\textbf
  {\bibinfo {volume} {37}},\ \bibinfo {pages} {2279} (\bibinfo {year}
  {1996})}\BibitemShut {NoStop}%
\bibitem [{\citenamefont {Vermant}\ \emph {et~al.}(1998)\citenamefont
  {Vermant}, \citenamefont {Walker}, \citenamefont {Moldenaers},\ and\
  \citenamefont {Mewis}}]{Mewis_OSP}%
  \BibitemOpen
  \bibfield  {author} {\bibinfo {author} {\bibfnamefont {J.}~\bibnamefont
  {Vermant}}, \bibinfo {author} {\bibfnamefont {P.}~\bibnamefont {Walker}},
  \bibinfo {author} {\bibfnamefont {P.}~\bibnamefont {Moldenaers}},\ and\
  \bibinfo {author} {\bibfnamefont {J.}~\bibnamefont {Mewis}},\ }\bibfield
  {title} {\enquote {\bibinfo {title} {{Orthogonal versus parallel
  superposition measurements.}}}\ }\href@noop {} {\bibfield  {journal}
  {\bibinfo  {journal} {J. Non-Newtonian Fluid Mech.}\ }\textbf {\bibinfo
  {volume} {79}},\ \bibinfo {pages} {173--189} (\bibinfo {year}
  {1998})}\BibitemShut {NoStop}%
\bibitem [{\citenamefont {Martinetti}, \citenamefont {Soulages},\ and\
  \citenamefont {Ewoldt}(2018)}]{LucaRHEMAOS_2018}%
  \BibitemOpen
  \bibfield  {author} {\bibinfo {author} {\bibfnamefont {L.}~\bibnamefont
  {Martinetti}}, \bibinfo {author} {\bibfnamefont {J.~M.}\ \bibnamefont
  {Soulages}},\ and\ \bibinfo {author} {\bibfnamefont {R.~H.}\ \bibnamefont
  {Ewoldt}},\ }\bibfield  {title} {\enquote {\bibinfo {title} {Continuous
  relaxation spectra for constitutive models in medium-amplitude oscillatory
  shear},}\ }\href@noop {} {\bibfield  {journal} {\bibinfo  {journal} {J.
  Rheol.}\ }\textbf {\bibinfo {volume} {62}},\ \bibinfo {pages} {1271--1298}
  (\bibinfo {year} {2018})}\BibitemShut {NoStop}%
\bibitem [{\citenamefont {Martinetti}\ and\ \citenamefont
  {Ewoldt}(2019)}]{LucaRHE_TSS2019}%
  \BibitemOpen
  \bibfield  {author} {\bibinfo {author} {\bibfnamefont {L.}~\bibnamefont
  {Martinetti}}\ and\ \bibinfo {author} {\bibfnamefont {R.~H.}\ \bibnamefont
  {Ewoldt}},\ }\bibfield  {title} {\enquote {\bibinfo {title} {Time-strain
  separability in medium amplitude oscillatory shear},}\ }\href@noop {}
  {\bibfield  {journal} {\bibinfo  {journal} {Phys. Fluids}\ }\textbf {\bibinfo
  {volume} {31}},\ \bibinfo {pages} {021213} (\bibinfo {year}
  {2019})}\BibitemShut {NoStop}%
\bibitem [{\citenamefont {Dullaert}\ and\ \citenamefont
  {Mewis}(2006)}]{DullaertMewis_structkinetics2006}%
  \BibitemOpen
  \bibfield  {author} {\bibinfo {author} {\bibfnamefont {K.}~\bibnamefont
  {Dullaert}}\ and\ \bibinfo {author} {\bibfnamefont {J.}~\bibnamefont
  {Mewis}},\ }\bibfield  {title} {\enquote {\bibinfo {title} {A structural
  kinetics model for thixotropy},}\ }\href@noop {} {\bibfield  {journal}
  {\bibinfo  {journal} {J. Non-Newtonian Fluid Mech.}\ }\textbf {\bibinfo
  {volume} {139}},\ \bibinfo {pages} {21--30} (\bibinfo {year}
  {2006})}\BibitemShut {NoStop}%
\bibitem [{\citenamefont {Wei}, \citenamefont {Solomon},\ and\ \citenamefont
  {Larson}(2018)}]{WeiSolomonLarson-JOR2018}%
  \BibitemOpen
  \bibfield  {author} {\bibinfo {author} {\bibfnamefont {Y.}~\bibnamefont
  {Wei}}, \bibinfo {author} {\bibfnamefont {M.~J.}\ \bibnamefont {Solomon}},\
  and\ \bibinfo {author} {\bibfnamefont {R.~G.}\ \bibnamefont {Larson}},\
  }\bibfield  {title} {\enquote {\bibinfo {title} {A multimode structural
  kinetics constitutive equation for the transient rheology of thixotropic
  elasto-viscoplastic fluids},}\ }\href@noop {} {\bibfield  {journal} {\bibinfo
   {journal} {J. Rheol.}\ }\textbf {\bibinfo {volume} {62}},\ \bibinfo {pages}
  {321--342} (\bibinfo {year} {2018})}\BibitemShut {NoStop}%
\bibitem [{\citenamefont {Johnston}(2006)}]{Johnston_SE-PRB2006}%
  \BibitemOpen
  \bibfield  {author} {\bibinfo {author} {\bibfnamefont {D.~C.}\ \bibnamefont
  {Johnston}},\ }\bibfield  {title} {\enquote {\bibinfo {title} {Stretched
  exponential relaxation arising from a continuous sum of exponential
  decays},}\ }\href@noop {} {\bibfield  {journal} {\bibinfo  {journal} {Phys.
  Rev. B}\ }\textbf {\bibinfo {volume} {74}},\ \bibinfo {pages} {184430}
  (\bibinfo {year} {2006})}\BibitemShut {NoStop}%
\bibitem [{\citenamefont {Berberan-Santos}, \citenamefont {Bodunov},\ and\
  \citenamefont {Valeur}(2005)}]{Santos2005}%
  \BibitemOpen
  \bibfield  {author} {\bibinfo {author} {\bibfnamefont {M.~N.}\ \bibnamefont
  {Berberan-Santos}}, \bibinfo {author} {\bibfnamefont {E.~N.}\ \bibnamefont
  {Bodunov}},\ and\ \bibinfo {author} {\bibfnamefont {B.}~\bibnamefont
  {Valeur}},\ }\bibfield  {title} {\enquote {\bibinfo {title} {{Mathematical
  functions for the analysis of luminescence decays with underlying
  distributions 1. Kohlrausch decay function (stretched exponential)}},}\
  }\href@noop {} {\bibfield  {journal} {\bibinfo  {journal} {J. Chem. Phys.}\
  }\textbf {\bibinfo {volume} {315}},\ \bibinfo {pages} {171--182} (\bibinfo
  {year} {2005})}\BibitemShut {NoStop}%
\bibitem [{\citenamefont {Ashby}(2011)}]{Ashby_book2011}%
  \BibitemOpen
  \bibfield  {author} {\bibinfo {author} {\bibfnamefont {M.~F.}\ \bibnamefont
  {Ashby}},\ }\href@noop {} {\emph {\bibinfo {title} {Materials Selection in
  Mechanical Design}}},\ \bibinfo {edition} {4th}\ ed.\ (\bibinfo  {publisher}
  {Elsevier},\ \bibinfo {year} {2011})\BibitemShut {NoStop}%
\bibitem [{\citenamefont {Bird}, \citenamefont {Armstrong},\ and\ \citenamefont
  {Hassager}(1987)}]{DPL_vol1}%
  \BibitemOpen
  \bibfield  {author} {\bibinfo {author} {\bibfnamefont {R.~B.}\ \bibnamefont
  {Bird}}, \bibinfo {author} {\bibfnamefont {R.~C.}\ \bibnamefont
  {Armstrong}},\ and\ \bibinfo {author} {\bibfnamefont {O.}~\bibnamefont
  {Hassager}},\ }\href@noop {} {\emph {\bibinfo {title} {Dynamics of Polymeric
  Liquids}}},\ Vol.~\bibinfo {volume} {1}\ (\bibinfo  {publisher} {Wiley},\
  \bibinfo {year} {1987})\BibitemShut {NoStop}%
\bibitem [{\citenamefont {Piau}(2007)}]{Piau_cpol2007}%
  \BibitemOpen
  \bibfield  {author} {\bibinfo {author} {\bibfnamefont {J.~M.}\ \bibnamefont
  {Piau}},\ }\bibfield  {title} {\enquote {\bibinfo {title} {{Carbopol gels:
  Elastoviscoplastic and slippery glasses made of individual swollen sponges
  Meso- and macroscopic properties, constitutive equations and scaling
  laws}},}\ }\href@noop {} {\bibfield  {journal} {\bibinfo  {journal} {J.
  Non-Newtonian Fluid Mech.}\ }\textbf {\bibinfo {volume} {144}},\ \bibinfo
  {pages} {1--29} (\bibinfo {year} {2007})}\BibitemShut {NoStop}%
\bibitem [{\citenamefont {Divoux}, \citenamefont {Grenard},\ and\ \citenamefont
  {Manneville}(2013)}]{Divoux_PRL2013}%
  \BibitemOpen
  \bibfield  {author} {\bibinfo {author} {\bibfnamefont {T.}~\bibnamefont
  {Divoux}}, \bibinfo {author} {\bibfnamefont {V.}~\bibnamefont {Grenard}},\
  and\ \bibinfo {author} {\bibfnamefont {S.}~\bibnamefont {Manneville}},\
  }\bibfield  {title} {\enquote {\bibinfo {title} {Rheological hysteresis in
  soft glassy materials},}\ }\href@noop {} {\bibfield  {journal} {\bibinfo
  {journal} {Phys. Rev. Lett.}\ }\textbf {\bibinfo {volume} {110}},\ \bibinfo
  {pages} {018304} (\bibinfo {year} {2013})}\BibitemShut {NoStop}%
\bibitem [{\citenamefont {Blackwell}\ \emph
  {et~al.}(2015{\natexlab{a}})\citenamefont {Blackwell}, \citenamefont
  {Deetjen}, \citenamefont {Gaudio},\ and\ \citenamefont
  {Ewoldt}}]{BCB_PhysFluids2015}%
  \BibitemOpen
  \bibfield  {author} {\bibinfo {author} {\bibfnamefont {B.~C.}\ \bibnamefont
  {Blackwell}}, \bibinfo {author} {\bibfnamefont {M.~E.}\ \bibnamefont
  {Deetjen}}, \bibinfo {author} {\bibfnamefont {J.~E.}\ \bibnamefont
  {Gaudio}},\ and\ \bibinfo {author} {\bibfnamefont {R.~H.}\ \bibnamefont
  {Ewoldt}},\ }\bibfield  {title} {\enquote {\bibinfo {title} {Sticking and
  splashing in yield-stress fluid drop impacts on coated surfaces},}\
  }\href@noop {} {\bibfield  {journal} {\bibinfo  {journal} {Phys. Fluids}\
  }\textbf {\bibinfo {volume} {27}},\ \bibinfo {pages} {043101} (\bibinfo
  {year} {2015}{\natexlab{a}})}\BibitemShut {NoStop}%
\bibitem [{\citenamefont {Sen}, \citenamefont {Morales},\ and\ \citenamefont
  {Ewoldt}(2020)}]{SSRHE_JFM2020}%
  \BibitemOpen
  \bibfield  {author} {\bibinfo {author} {\bibfnamefont {S.}~\bibnamefont
  {Sen}}, \bibinfo {author} {\bibfnamefont {A.~G.}\ \bibnamefont {Morales}},\
  and\ \bibinfo {author} {\bibfnamefont {R.~H.}\ \bibnamefont {Ewoldt}},\
  }\bibfield  {title} {\enquote {\bibinfo {title} {Viscoplastic drop impact on
  thin films},}\ }\href@noop {} {\bibfield  {journal} {\bibinfo  {journal} {J.
  Fluid Mech.}\ }\textbf {\bibinfo {volume} {891}},\ \bibinfo {pages} {A27}
  (\bibinfo {year} {2020})}\BibitemShut {NoStop}%
\bibitem [{\citenamefont {Armstrong}\ \emph {et~al.}(2016)\citenamefont
  {Armstrong}, \citenamefont {Beris}, \citenamefont {Rogers},\ and\
  \citenamefont {Wagner}}]{ArmWag_JOR2016}%
  \BibitemOpen
  \bibfield  {author} {\bibinfo {author} {\bibfnamefont {M.~J.}\ \bibnamefont
  {Armstrong}}, \bibinfo {author} {\bibfnamefont {A.~N.}\ \bibnamefont
  {Beris}}, \bibinfo {author} {\bibfnamefont {S.~A.}\ \bibnamefont {Rogers}},\
  and\ \bibinfo {author} {\bibfnamefont {S.~A.}\ \bibnamefont {Wagner}},\
  }\bibfield  {title} {\enquote {\bibinfo {title} {Dynamic shear rheology of a
  thixotropic suspension: Comparison of an improved structure-based model with
  large amplitude oscillatory shear experiments},}\ }\href@noop {} {\bibfield
  {journal} {\bibinfo  {journal} {J. Rheol.}\ }\textbf {\bibinfo {volume}
  {60}},\ \bibinfo {pages} {433--450} (\bibinfo {year} {2016})}\BibitemShut
  {NoStop}%
\bibitem [{\citenamefont {Ewoldt}, \citenamefont {Johnston},\ and\
  \citenamefont {Caretta}(2015)}]{RHE_baddata}%
  \BibitemOpen
  \bibfield  {author} {\bibinfo {author} {\bibfnamefont {R.~H.}\ \bibnamefont
  {Ewoldt}}, \bibinfo {author} {\bibfnamefont {M.~T.}\ \bibnamefont
  {Johnston}},\ and\ \bibinfo {author} {\bibfnamefont {L.~M.}\ \bibnamefont
  {Caretta}},\ }\bibfield  {title} {\enquote {\bibinfo {title} {Experimental
  challenges of shear rheology: how to avoid bad data},}\ }in\ \href@noop {}
  {\emph {\bibinfo {booktitle} {Complex Fluids in Biological Systems}}},\
  \bibinfo {editor} {edited by\ \bibinfo {editor} {\bibfnamefont
  {S.}~\bibnamefont {Spagnolie}}}\ (\bibinfo  {publisher} {Springer Biological
  Engineering Series},\ \bibinfo {year} {2015})\ pp.\ \bibinfo {pages}
  {207--241}\BibitemShut {NoStop}%
\bibitem [{\citenamefont {Singh}, \citenamefont {Soulages},\ and\ \citenamefont
  {Ewoldt}(2019)}]{Singh_2019}%
  \BibitemOpen
  \bibfield  {author} {\bibinfo {author} {\bibfnamefont {P.~K.}\ \bibnamefont
  {Singh}}, \bibinfo {author} {\bibfnamefont {J.~M.}\ \bibnamefont
  {Soulages}},\ and\ \bibinfo {author} {\bibfnamefont {R.~H.}\ \bibnamefont
  {Ewoldt}},\ }\bibfield  {title} {\enquote {\bibinfo {title} {On fitting data
  for parameter estimates: residual weighting and data representation},}\
  }\href@noop {} {\bibfield  {journal} {\bibinfo  {journal} {Rheol. Acta}\
  }\textbf {\bibinfo {volume} {58}},\ \bibinfo {pages} {341--359} (\bibinfo
  {year} {2019})}\BibitemShut {NoStop}%
\bibitem [{\citenamefont {Tikhonov}\ and\ \citenamefont
  {Arsenin}(1979)}]{Tikhonov_1979}%
  \BibitemOpen
  \bibfield  {author} {\bibinfo {author} {\bibfnamefont {A.~N.}\ \bibnamefont
  {Tikhonov}}\ and\ \bibinfo {author} {\bibfnamefont {V.~Y.}\ \bibnamefont
  {Arsenin}},\ }\bibfield  {title} {\enquote {\bibinfo {title} {Solutions of
  ill-posed problems},}\ }\href@noop {} {\bibfield  {journal} {\bibinfo
  {journal} {SIAM Rev.}\ }\textbf {\bibinfo {volume} {21}},\ \bibinfo {pages}
  {266--267} (\bibinfo {year} {1979})}\BibitemShut {NoStop}%
\bibitem [{\citenamefont {Kontogiorgos}(2010{\natexlab{a}})}]{Kont_2010}%
  \BibitemOpen
  \bibfield  {author} {\bibinfo {author} {\bibfnamefont {V.}~\bibnamefont
  {Kontogiorgos}},\ }\bibfield  {title} {\enquote {\bibinfo {title}
  {{Calculation of relaxation spectra from mechanical spectra in MATLAB}},}\
  }\href@noop {} {\bibfield  {journal} {\bibinfo  {journal} {Polym. Test.}\
  }\textbf {\bibinfo {volume} {29}},\ \bibinfo {pages} {1021--1025} (\bibinfo
  {year} {2010}{\natexlab{a}})}\BibitemShut {NoStop}%
\bibitem [{\citenamefont {Kontogiorgos}(2010{\natexlab{b}})}]{Kontogiorgos}%
  \BibitemOpen
  \bibfield  {author} {\bibinfo {author} {\bibfnamefont {V.}~\bibnamefont
  {Kontogiorgos}},\ }\bibfield  {title} {\enquote {\bibinfo {title}
  {Calculation of relaxation spectra from stress relaxation measurements},}\
  }in\ \href@noop {} {\emph {\bibinfo {booktitle} {Biopolymers}}},\ \bibinfo
  {editor} {edited by\ \bibinfo {editor} {\bibfnamefont {M.}~\bibnamefont
  {Elnashar}}}\ (\bibinfo  {publisher} {IntechOpen},\ \bibinfo {address}
  {Rijeka},\ \bibinfo {year} {2010})\ Chap.~\bibinfo {chapter} {25}\BibitemShut
  {NoStop}%
\bibitem [{PCH()}]{PCH}%
  \BibitemOpen
  \href@noop {} {\enquote {\bibinfo {title} {{MATLAB regtools by Per Christian
  Hansen}},}\ }\bibinfo {howpublished}
  {\url{https://www.mathworks.com/matlabcentral/fileexchange/52-regtools}}\BibitemShut
  {NoStop}%
\bibitem [{\citenamefont {Goodeve}\ and\ \citenamefont
  {Whitfield}(1938)}]{Goodeve1938}%
  \BibitemOpen
  \bibfield  {author} {\bibinfo {author} {\bibfnamefont {C.}~\bibnamefont
  {Goodeve}}\ and\ \bibinfo {author} {\bibfnamefont {G.}~\bibnamefont
  {Whitfield}},\ }\bibfield  {title} {\enquote {\bibinfo {title} {The
  measurement of thixotropy in absolute units},}\ }\href@noop {} {\bibfield
  {journal} {\bibinfo  {journal} {Trans. Faraday Soc.}\ }\textbf {\bibinfo
  {volume} {34}},\ \bibinfo {pages} {511--520} (\bibinfo {year}
  {1938})}\BibitemShut {NoStop}%
\bibitem [{\citenamefont {Moore}(1959)}]{Moore1959}%
  \BibitemOpen
  \bibfield  {author} {\bibinfo {author} {\bibfnamefont {F.}~\bibnamefont
  {Moore}},\ }\bibfield  {title} {\enquote {\bibinfo {title} {The rheology of
  ceramic slips and bodies},}\ }\href@noop {} {\bibfield  {journal} {\bibinfo
  {journal} {Trans. Br. Ceram. Soc.}\ }\textbf {\bibinfo {volume} {58}},\
  \bibinfo {pages} {470--494} (\bibinfo {year} {1959})}\BibitemShut {NoStop}%
\bibitem [{\citenamefont {Sen}, \citenamefont {Morales},\ and\ \citenamefont
  {Ewoldt}(2021)}]{SSRHE_PRF2021}%
  \BibitemOpen
  \bibfield  {author} {\bibinfo {author} {\bibfnamefont {S.}~\bibnamefont
  {Sen}}, \bibinfo {author} {\bibfnamefont {A.~G.}\ \bibnamefont {Morales}},\
  and\ \bibinfo {author} {\bibfnamefont {R.~H.}\ \bibnamefont {Ewoldt}},\
  }\bibfield  {title} {\enquote {\bibinfo {title} {Thixotropy in viscoplastic
  drop impact on thin films},}\ }\href@noop {} {\bibfield  {journal} {\bibinfo
  {journal} {Phys. Rev. Fluids}\ }\textbf {\bibinfo {volume} {6}},\ \bibinfo
  {pages} {043301} (\bibinfo {year} {2021})}\BibitemShut {NoStop}%
\bibitem [{\citenamefont {Jamali}, \citenamefont {Armstrong},\ and\
  \citenamefont {McKinley}(2019)}]{Jamali_PRL2019}%
  \BibitemOpen
  \bibfield  {author} {\bibinfo {author} {\bibfnamefont {S.}~\bibnamefont
  {Jamali}}, \bibinfo {author} {\bibfnamefont {R.~C.}\ \bibnamefont
  {Armstrong}},\ and\ \bibinfo {author} {\bibfnamefont {G.~H.}\ \bibnamefont
  {McKinley}},\ }\bibfield  {title} {\enquote {\bibinfo {title} {Multiscale
  nature of thixotropy and rheological hysteresis in attractive colloidal
  suspensions under shear},}\ }\href@noop {} {\bibfield  {journal} {\bibinfo
  {journal} {Phys. Rev. Lett.}\ }\textbf {\bibinfo {volume} {123}},\ \bibinfo
  {pages} {248003} (\bibinfo {year} {2019})}\BibitemShut {NoStop}%
\bibitem [{\citenamefont {Anscombe}(1973)}]{Anscombe}%
  \BibitemOpen
  \bibfield  {author} {\bibinfo {author} {\bibfnamefont {F.~J.}\ \bibnamefont
  {Anscombe}},\ }\bibfield  {title} {\enquote {\bibinfo {title} {Graphs in
  statistical analysis},}\ }\href@noop {} {\bibfield  {journal} {\bibinfo
  {journal} {Am. Stat.}\ }\textbf {\bibinfo {volume} {27}},\ \bibinfo {pages}
  {17--21} (\bibinfo {year} {1973})}\BibitemShut {NoStop}%
\bibitem [{\citenamefont {Freund}\ and\ \citenamefont
  {Ewoldt}(2015)}]{FreundEwoldt}%
  \BibitemOpen
  \bibfield  {author} {\bibinfo {author} {\bibfnamefont {J.~B.}\ \bibnamefont
  {Freund}}\ and\ \bibinfo {author} {\bibfnamefont {R.~H.}\ \bibnamefont
  {Ewoldt}},\ }\bibfield  {title} {\enquote {\bibinfo {title} {{Quantitative
  rheological model selection: good fits versus credible models using Bayesian
  inference}},}\ }\href@noop {} {\bibfield  {journal} {\bibinfo  {journal} {J.
  Rheol.}\ }\textbf {\bibinfo {volume} {59}},\ \bibinfo {pages} {667--701}
  (\bibinfo {year} {2015})}\BibitemShut {NoStop}%
\bibitem [{\citenamefont {Mours}\ and\ \citenamefont
  {Winter}(1994)}]{Winter_1994}%
  \BibitemOpen
  \bibfield  {author} {\bibinfo {author} {\bibfnamefont {M.}~\bibnamefont
  {Mours}}\ and\ \bibinfo {author} {\bibfnamefont {H.~H.}\ \bibnamefont
  {Winter}},\ }\bibfield  {title} {\enquote {\bibinfo {title} {Time-resolved
  rheometry},}\ }\href@noop {} {\bibfield  {journal} {\bibinfo  {journal}
  {Rheol. Acta}\ }\textbf {\bibinfo {volume} {33}},\ \bibinfo {pages}
  {385--397} (\bibinfo {year} {1994})}\BibitemShut {NoStop}%
\bibitem [{\citenamefont {Laukkanen}, \citenamefont {Winter},\ and\
  \citenamefont {Sepp\"{a}l\"{a}}(2018)}]{Winter_2018}%
  \BibitemOpen
  \bibfield  {author} {\bibinfo {author} {\bibfnamefont {O.-V.}\ \bibnamefont
  {Laukkanen}}, \bibinfo {author} {\bibfnamefont {H.~H.}\ \bibnamefont
  {Winter}},\ and\ \bibinfo {author} {\bibfnamefont {J.}~\bibnamefont
  {Sepp\"{a}l\"{a}}},\ }\bibfield  {title} {\enquote {\bibinfo {title}
  {Characterization of physical aging by time-resolved rheometry: fundamentals
  and application to bituminous binders},}\ }\href@noop {} {\bibfield
  {journal} {\bibinfo  {journal} {Rheol. Acta}\ }\textbf {\bibinfo {volume}
  {57}},\ \bibinfo {pages} {745--756} (\bibinfo {year} {2018})}\BibitemShut
  {NoStop}%
\bibitem [{\citenamefont {Geri}\ \emph {et~al.}(2018)\citenamefont {Geri},
  \citenamefont {Keshavarz}, \citenamefont {Divoux}, \citenamefont {Clasen},
  \citenamefont {Curtis},\ and\ \citenamefont {McKinley}}]{GeriGHM_PRX2018}%
  \BibitemOpen
  \bibfield  {author} {\bibinfo {author} {\bibfnamefont {M.}~\bibnamefont
  {Geri}}, \bibinfo {author} {\bibfnamefont {B.}~\bibnamefont {Keshavarz}},
  \bibinfo {author} {\bibfnamefont {T.}~\bibnamefont {Divoux}}, \bibinfo
  {author} {\bibfnamefont {C.}~\bibnamefont {Clasen}}, \bibinfo {author}
  {\bibfnamefont {D.~J.}\ \bibnamefont {Curtis}},\ and\ \bibinfo {author}
  {\bibfnamefont {G.~H.}\ \bibnamefont {McKinley}},\ }\bibfield  {title}
  {\enquote {\bibinfo {title} {Time-resolved mechanical spectroscopy of soft
  materials via optimally windowed chirps},}\ }\href@noop {} {\bibfield
  {journal} {\bibinfo  {journal} {Phys. Rev. X}\ }\textbf {\bibinfo {volume}
  {8}},\ \bibinfo {pages} {041042} (\bibinfo {year} {2018})}\BibitemShut
  {NoStop}%
\bibitem [{\citenamefont {Ewoldt}\ and\ \citenamefont
  {Saengow}(2022)}]{ChaiRHE_ARFM}%
  \BibitemOpen
  \bibfield  {author} {\bibinfo {author} {\bibfnamefont {R.~H.}\ \bibnamefont
  {Ewoldt}}\ and\ \bibinfo {author} {\bibfnamefont {C.}~\bibnamefont
  {Saengow}},\ }\bibfield  {title} {\enquote {\bibinfo {title} {Designing
  complex fluids},}\ }\href@noop {} {\bibfield  {journal} {\bibinfo  {journal}
  {Annu. Rev. Fluid Mech.}\ }\textbf {\bibinfo {volume} {54}},\ \bibinfo
  {pages} {413--441} (\bibinfo {year} {2022})}\BibitemShut {NoStop}%
\bibitem [{\citenamefont {Corker}\ \emph {et~al.}(2019)\citenamefont {Corker},
  \citenamefont {Ng}, \citenamefont {Poole},\ and\ \citenamefont
  {Garc\'ia-Tu\~n\'on}}]{Poole_SM2019}%
  \BibitemOpen
  \bibfield  {author} {\bibinfo {author} {\bibfnamefont {A.}~\bibnamefont
  {Corker}}, \bibinfo {author} {\bibfnamefont {H.~C.-H.}\ \bibnamefont {Ng}},
  \bibinfo {author} {\bibfnamefont {R.~J.}\ \bibnamefont {Poole}},\ and\
  \bibinfo {author} {\bibfnamefont {E.}~\bibnamefont {Garc\'ia-Tu\~n\'on}},\
  }\bibfield  {title} {\enquote {\bibinfo {title} {{3D printing with 2D
  colloids: designing rheology protocols to predict `printability' of
  soft-materials}},}\ }\href@noop {} {\bibfield  {journal} {\bibinfo  {journal}
  {Soft Matter}\ }\textbf {\bibinfo {volume} {15}},\ \bibinfo {pages}
  {1444--1456} (\bibinfo {year} {2019})}\BibitemShut {NoStop}%
\bibitem [{\citenamefont {Blackwell}\ \emph
  {et~al.}(2015{\natexlab{b}})\citenamefont {Blackwell}, \citenamefont
  {Nadhan}, \citenamefont {Wu},\ and\ \citenamefont {Ewoldt}}]{BCB_GFM2015}%
  \BibitemOpen
  \bibfield  {author} {\bibinfo {author} {\bibfnamefont {B.~C.}\ \bibnamefont
  {Blackwell}}, \bibinfo {author} {\bibfnamefont {A.~E.}\ \bibnamefont
  {Nadhan}}, \bibinfo {author} {\bibfnamefont {A.}~\bibnamefont {Wu}},\ and\
  \bibinfo {author} {\bibfnamefont {R.~H.}\ \bibnamefont {Ewoldt}},\
  }\href@noop {} {\enquote {\bibinfo {title} {{Fire suppression with
  yield-stress fluids}},}\ }\bibinfo {howpublished}
  {\href{https://doi.org/10.1103/APS.DFD.2015.GFM.V0007}{Gallery of Fluid
  Motion, APS Division of Fluid Dynamics 68th Annual Meeting}} (\bibinfo {year}
  {2015}{\natexlab{b}})\BibitemShut {NoStop}%
\bibitem [{\citenamefont {Sen}\ and\ \citenamefont
  {Ewoldt}(2018)}]{SSRHE_GFM2018}%
  \BibitemOpen
  \bibfield  {author} {\bibinfo {author} {\bibfnamefont {S.}~\bibnamefont
  {Sen}}\ and\ \bibinfo {author} {\bibfnamefont {R.~H.}\ \bibnamefont
  {Ewoldt}},\ }\href@noop {} {\enquote {\bibinfo {title} {{Stick or Splash?
  Thixotropy Decides!}}}\ }\bibinfo {howpublished}
  {\href{https://doi.org/10.1103/APS.DFD.2018.GFM.V0036}{Gallery of Fluid
  Motion, APS Division of Fluid Dynamics 71st Annual Meeting}} (\bibinfo {year}
  {2018})\BibitemShut {NoStop}%
\end{thebibliography}%

\end{document}